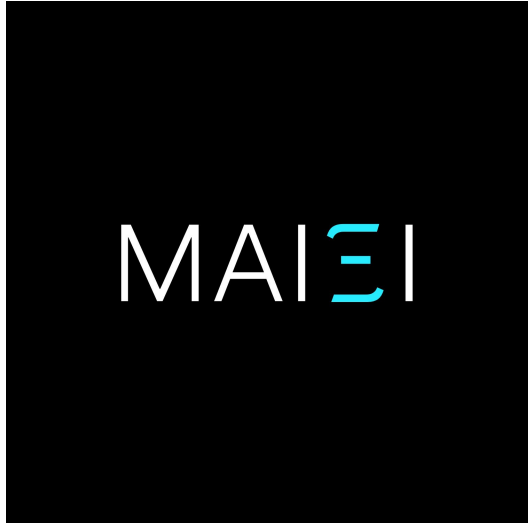

Montreal AI Ethics Institute

https://montrealethics.ai

**Prepared by:**

Mirka Snyder Caron, Sr. Associate, MAIEI
Abhishek Gupta, Founder, MAIEI and ML Engineer, Microsoft

**Response to Office of the Privacy Commissioner of Canada Consultation Proposals pertaining to amendments to PIPEDA relative to Artificial Intelligence.**

**Date: March 2, 2020**

**Introductive Summary:**

In February 2020, the Montreal AI Ethics Institute **(MAIEI)** was invited by the Office of the Privacy Commissioner of Canada **(OPCC)** to provide for comments both at a closed roundtable and in writing on the OPCC consultation proposal for amendments relative to Artificial Intelligence **(AI)**, to the Canadian privacy legislation, the *Personal Information Protection and Electronic Documents Act* **(PIPEDA)**.

The present document includes MAIEI comments and recommendations in writing. Per MAIEI's mission and mandate to act as a catalyst for public feedback pertaining to AI Ethics and regulatory technology developments, as well as to provide for public competence-building workshops on critical topics in such domains, the reader will also find such public feedback and propositions by Montrealers who participated at MAIEI's workshops, submitted as Schedule 1 to the present report. For each of OPCC 12 proposals, and underlying questions, as described on its [website](#), MAIEI provides a short reply, a summary list of recommendations, as well as comments relevant to the question at hand.

We leave you with three general statements to keep in mind while going through the next pages:

1) **AI systems should be used to augment human capacity for meaningful and purposeful connections and associations, not as a substitute for trust.**

2) **Humans have collectively accepted to uphold the rule of law, but for machines, the code is rule. Where socio-technical systems are deployed to make important decisions, profiles or inferences about individuals, we will increasingly have to attempt the difficult exercise of drafting and encoding our law in a manner learnable by machines.**

3) **Let us work collectively towards a world where Responsible AI becomes the rule, before our socio-technical systems become "too connected to fail"[1].**

Best,
**The Montreal AI Ethics Institute[2]**

---

[1] ARNER, Douglas, OTHERS, *The Dark Side of Digital Financial Transformation: The New Risks of Fintech and the Rise of TechRisk*, University of Hong Kong Faculty of Law Research Paper No. 2019/112, 37 pages, p. 5.
[2] https://montrealethics.ai

**Proposal 1: Incorporate a definition of AI within the law that would serve to clarify which legal rules would apply only to it, while other rules would apply to all processing, including AI**

1. **Should AI be governed by the same rules as other forms of processing, potentially enhanced as recommended in this paper (which means there would be no need for a definition and the principles of technological neutrality would be preserved) or should certain rules be limited to AI due to its specific risks to privacy and, consequently, to other human rights?**

<u>Short Reply</u>:

Maintain the general legal framework with the principle of technological neutrality, but provide prescribed specific and practical regulation or hard law Code of Practice for today's emerging technologies.

<u>List of Recommendations:</u>

1. **Maintain technologically neutral framework of the general law.**

2. **Amend the general law as to show explicit awareness and application to technological developments, while associating such sections to prescribed compliance frames within the supporting Regulation.**

3. **Add particular prescription pertaining to existing technological developments in Regulation or prescribe Code of Practice (not soft law).**

4. **Cooperate with key foreign jurisdiction's relevant and authorized representatives to discuss adequacy levels.**

5. **Design and cooperate with human rights regulators, representatives and tribunals to bridge potential gaps and avoid regulatory overlap.**

6. **Establish multidisciplinary teams to design prescribed Regulation in both a human-readable and machine-enforceable format.**

**Comments:**

We consider the optimal approach to be to both maintain the technological neutrality approach within the general framework of the law, while however adding specifically curtailed sections on critical applications of technology, as aligned with present and existing developments.

The purpose of these specificities would be to ensure necessary guidance for businesses to frame their business and legal risks in a more certain manner while ensuring sufficient, appropriate and concrete protection of individuals.

Otherwise, the ambiguity and uncertainty relating to what is to be reasonably expected or reasonably interpreted as appropriate security safeguards in the neutral framework may increase both societal costs, as well as legal provisions and operational business costs pertaining to managing compliance risks. This may correlatively weaken or counter the benefits of the neutral approach.

From a technical perspective, maintenance of technical neutrality may be hard given that there are many capabilities of AI systems that evade regulation under the current law. Instead, utilizing a domain based approach, in other words, providing for specific legislation for a particular industry or field, is better given the vastly different paces of development in different subfields of AI and their subsequent capabilities and techniques involving use of data. Folding legislation into industry specific laws will yield more meaningful control compared to generic principles which can be followed to the letter yet evade the spirit of the law.

This approach does not otherwise contradict the technological neutrality principle. It augments its applicability, showing explicit awareness, knowledge of such technologies, as well as it provides practical guidance about ongoing business and technological developments for and to key business players. It also shows better promise in maintaining sufficiently broad applicability to other human and technical measures or tools. This will enable both appropriate protection and operational and compliance clarity to businesses concerned.

We consider specific articles pertaining to existing developments should be prescribed within regulation, and less so in the general framework of the law. This will enable faster response in terms of amending regulation as compared to the law. It also better maintains the technological neutrality principle of the general law.

Due to the cross-border applicability of existing technologies and flow of data, careful consideration should be given to ensuring the Canadian privacy framework meets similar or essentially similar adequacy levels of security safeguards with key international trade jurisdictions, where appropriate (eg. EU Commission decision-making of adequacy per *General Data Protection Regulation*[3] **(GDPR)**).

We also recommend that, aside from privacy experts and regulators, such representatives of human rights policy-making groups, and human rights tribunals, be made part of the consultation and design process, to both avoid overlap and potential jurisdictional competency and administrative mandate issues, as well as to avoid gaps and deficiencies between legislated activities and fields.

Finally, we recommend the drafting of the prescribed specific sections within regulation be designed and drafted within a multidisciplinary team, in conjunction with policymakers, lawyers and technical and business staff involved in the design, development, deployment and maintenance of AI systems, to format the drafting as close as possible to what could be effectively programmed within existing software platforms and systems. The purpose would be to facilitate privacy by default through coded techniques. This would mitigate the loss in translation risk from the conversion of human-readable legal language to machine-enforceable coded rules.

**2. If certain rules should apply to AI only, how should AI be defined in the law to help clarify the application of such rules?**

**Short Reply:**

The attempt of defining AI at law may fall short of the potentially limitless opportunities for associations of databases, algorithms, and other technologies or tools which are still emerging to date.

Perhaps attempting to define "systems" as an "associated network or ecosystem of software platforms, databases, technologies and other tools", which would include machine-learning algorithmic techniques, irrespective of whether they have been automated or not, would be more appropriate.

---

[3]*General Data Protection Regulation*, (EU) 2016/679.

**List of Recommendations:**

1. **Provide sense of security to relevant individuals and clear guidance to businesses concerned through contextual guidelines or preamble explicitly referring to AI, but the latter to be interpreted and perceived as a narrow concept falling into a broader protective scope of application which include any and all other potential human, technical, and technological associations, tools, or measures that may be designed, developed or scaled.**

2. **Provide for prescribed regulation for such presently existing associations and systems, to be reviewed and modified as required based on ongoing technological developments.**

3. **Promote and educate the public at large to convert the "hype" of the word "AI" to the much more representative concept of "socio-technical systems", which would encompass the concept of "algorithms", to make it clear that there is an element of human interaction which may lead to certain consequences.**

**Comments:**

It is extremely important to note that AI is a buzzword which has fed the imagination and attention of the public and fuelled massive investment strategies since 1942[4]. Depending on the level of sophistication of different members of the public or markets, it has been associated with different promises, different purposes, varying ethical, civilian, military, or malicious objectives, varying defined scopes or application, as well as differing levels of autonomy depending on the types of models used, and finally, even socially perceived as having varying levels of intelligence akin to sentient beings.

We have seen different bills, legislation, standard-setting bodies and other regulatory entities attempt to define an appropriate scope in defining AI systems, or algorithms.

---

[4]NORVIG, Peter, "Artificial Intelligence: A Modern Approach", Pearson, 3rd edition, 2016, p. 18ss.

The following are relevant excerpts of a published work:

**"Definitions of AI**
Algorithms programmed today are more flexible than traditional statistical
models. They can self-determine pattern, identification, recognition, and categorization through learning strategies, reaching programmed goals without prior guidelines. For instance, they can apply a probabilistic approach to analyzing the semantics for natural language processing (NLP) purposes by going through significant amounts of data. There is growing implementation of ''narrow AI,'' through which algorithms already provide impressive predictive calculation capabilities, but with very limited, task-specific applications; for example, Google's DeepMind AlphaGo[5] self-learned the game of Go and eventually beat the world champion in 2017.

However, there have been no significant breakthroughs in "general AI,"[6] which would involve cognitive-like capabilities independently juggling different datasets, ideas, and tasks without single-application limitations.

Proposed definitions for AI differ greatly due to varying levels of understanding of the technology among laypersons, ''acqui-hiring'' investors, startups, and the marketing teams of private corporations, as well as in public policy. For instance, providing a natural language interpretation, the Merriam-Webster dictionary defines AI as ''a branch of computer science dealing with the simulation of intelligent behavior in computers'' or ''the capability of a machine to imitate intelligent human behavior.''

Similarly, a US bill, the FUTURE of Artificial Intelligence Act (''FUTURE Act''), provides five different legal definitions for AI, which can be summarized as follows:

1. a task-performing artificial system functioning in ''unpredictable circumstances, without significant human oversight,'' that can ''learn from experience'' and ''improve [its] performance . . .''

---

[5]Deepmind, AlphaGo, https://deepmind.com/research/case-studies/alphago-the-story-so-far
[6]VINCENT, James, "This is when experts think we'll build a truly intelligent AI", https://www.theverge.com/2018/11/27/18114362/ai-artificial-general-intelligence-when-achieved-martin-ford-book

2. a system that can ''think like humans, such as cognitive architectures and neural networks'';

3. a system that can ''act like humans" by passing the Turing test or other test through NLP or learning;

4. ''a set of techniques, including ML,'' intended to ''approximate'' a ''cognitive task''; and
5. a goal-oriented system acting like a rational agent or with a robotic body.

Finally, the Financial Stability Board (FSB)[7] also published definitions applicable to the banking industry per its role as the supervising coordinator of financial standard-setting bodies:

**Artificial intelligence:** ''Application of computational tools to address tasks traditionally requiring human sophistication,'' or ''Theory and development of computer systems able to perform tasks that traditionally required human intelligence.''

**Machine learning:** ''Method of designing a sequence of actions to solve a problem, known as algorithms, which optimize automatically through experience and with limited or no human intervention. The algorithm may find patterns in large amounts of data from increasingly diverse and innovative sources.''

In addition to the lack of definitional precision, there is presently much hype involved in using buzzwords like ''AI'' or ''artificial intelligence,'' a term coined by John McCarthy in 1956. It is therefore difficult to provide an accurate or globally consented upon definition of what constitutes or is perceived as AI. The EU discussion paper on the proposal for the Civil Law Rules on Robotics, particularly regarding imminent voting on ''electronic persons'' liability, is evidence of such debated attempts to predict future AI developments and define an acceptable regulatory scope."[8]

In addition, the OECD definition of AI[9] appears to us to fall short in terms of providing a complete definition in cases where the objective function of a system is noy alway human defined. As an example, a reinforcement-learning system may have been tasked to not lose a game, and in such effect, may learn to pause the game[10] just before losing, whereas a human defined objective would have been to win the game. There can be sub-goals that emerge under the human defined objectives which can distort how the high-level goal is reached, which may result in the AI following the objective in letter but not in spirit. This problem is more fully addressed as a case in AI safety and there are numerous examples in toy domains that illustrate the shortcoming of such a definition where there can be emergent behaviour as a function of the complexity of the system.

---

[7]FINANCIAL STABILITY BOARD, "Artificial Intelligence and Machine Learning in Financial Services: Market Developments and Financial Stability Implications",
https://www.fsb.org/wp-content/uploads/P011117.pdf
[8]SNYDER CARON, Mirka, "The Transformative Effect of AI on the Banking Industry", Banking & Finance Law Review, April 2019, 34 BFLR 79-345
[9]OECD Legal Instruments, https://legalinstruments.oecd.org/en/instruments/OECD-LEGAL-0449

The fundamental concern that is not captured in any of the definitions provided are that AI itself is an evolving field. Over time, things that were initially considered AI are no longer viewed as such because the systems fade to the background. For example, predictive typing on the phone, and perhaps soon predictive typing in electronic messaging platforms, conversation nudging by voice assistants, automatic parsing and calendar population from emails, etc.

It is also extremely important to note that all such programmable techniques and algorithms called AI, whether with or without machine-learning, have no or very minimal risk of triggering adverse consequences if they are not connected to databases, or other tools. Without input of data, these algorithms are effectively rendered useless and harmless. **It is the "connection" or "association"** of such algorithms to the databases and sensory data-tools (eg. camera), which brings about the risk of adverse consequences.

As such, we are of the position that attempting to legally define the **potentially limitless and easily scalable "package"** of techniques and algorithms, either embedded within existing software or systems, or added-on and then connected through Application Programming Interfaces **(APIs)** or other techniques to a potentially limitless amount of databases, which are then connected and associated within different tools, such as cameras, other software, or other robotics, will inevitably fall short from effectively dealing with the existing developments, or effectively mitigating adverse consequences.

The exercise of defining a concept means circumscribing the limits of such specific concept, whereas the disruptive nature of present developments that we see is due to:

- the potentially limitless volume of connections and associations made by human business developers and programmers; between

---

[10] METZ, Cade, "Forget doomsday AI Google worried housekeeping bots gone bad", https://www.wired.com/2016/06/forget-doomsday-ai-google-worried-housekeeping-bots-gone-bad/

- a potentially limitless amount of coding languages, software systems, mathematical techniques and algorithms; with
- a potentially limitless number of other technologies and robotics, for
- a potentially limitless number of social, artistic, academic or business useful, and less useful, applications, purposes and functions, with
- a potentially limitless amount of natural and human data.

In the end, limiting the potential of such associations in an abstract legal concept may create unwanted or unforeseen externalities which will fall outside the scope of legal application, and may understate the importance of other future potential associations that will be designed in the short to longer term.

We have seen many NGOs, standard-setting bodies, policymakers and other key players attempt to provide a definition for what a given jurisdiction or corporation considers as AI, and all such definitions appear to show some form of universal foundation, but with varying scopes of applications and functions. Defining it will simply add to the growing amount of unsatisfying definitions attempting to conceptualize the disruptive nature of this package.

The most important advantage we see in attempting to define AI explicitly in a legislation is the sense of security and trust this may provide to individuals and citizens concerned, by the legislator and regulators, as they show that they are aware of the existing technological developments, and are taking measures for which private key players may not attempt to abusively take advantage of individuals for their personal or business gains, while at the same time assisting such businesses in mitigating compliance and business risks, to orient their business, investment and costs strategies in a manner that is aligned with public and regulators' expectations.

Based on our experience, public fears surrounding AI can be summarized as threefold and focus on the sense of powerlessness of the individual:

1. Being or feeling **powerless** in the perception, real or imagined, that **AI may control or harm them, and impose** unwanted judgments, false or negative inferences about themselves, restrictions to their behaviour, survival and social needs, beliefs, conduct, or identity (**"Insecurity linked to potential emergence of a threat"**)
2. Being or feeling **powerless** in the perception, real or imagined, that **other humans** providing them with critical products or services, **may control or harm them, or control access to, and impose** unwanted judgments, false or negative inferences about themselves, restrictions to their behaviour, survival and social needs, beliefs, conduct or identity **("Insecurity linked to exacerbating and consolidating social hierarchies, as well as control and power in the hands of a selfish few over many")**

3. Being and feeling **powerless** in the perception, real or imagined, of being **unable to sufficiently adapt** to social and business changes in due time, as well as **being forced to change** some of their behaviours or methods, **("Insecurity in the perception of a threat, real or imagined, to their sense of self, identity, and survival")**

The advantage of reassuring the public is certainly not to be understated, since effectively assuaging individuals that their human rights and freedoms will be prioritized and secured while providing additional clarity as to the manner in which businesses in Canada are expected to behave in respective markets, may ensure social and economic stability and interest, while showing support and clear guidance for innovation.

As such, we suggest that the optimal approach would be to provide for explicit wording in the general framework within preamble or through contextual guidelines **referring to AI, but within a broader context of technological application.** In other words, provide drafting in a manner that it would be perceived by reader and individuals as to provide **broader protection than that specific to AI,** and to encompass both and all human and technical or machine techniques and measures.

Furthermore, adding explicit language linking objectives of the law as to broader scope than that of the right of privacy alone, and more as all-encompassing of all human rights and freedoms, would be optimal, similarly to that of the preamble in the GDPR. This would evidently be subject to limits of competency and mandate, as this may redefine the scope of application of the privacy law in Canada, but enhancing it into a human rights law framework in a technological and digital context, in a less siloed perspective, would be recommended, although, greater cooperation with human rights tribunals will be required.

The above may provide for contextual or historical reading of existing technological developments, but will also provide for a more general framework. It will also avoid having businesses attempt to exclude the applicability of certain obligations to other techniques or technological associations which would otherwise appear or would be reasonably interpreted as being included within scope. The drafting of such may be reviewed and modified through the 5-yr review clause, for the legislation, and in a shorter span on time for regulation, and can be aligned with newly enhanced technical associations or applications.

**Proposal 2: Adopt a rights-based approach in the law, whereby data protection principles are implemented as a means to protect a broader right to privacy—recognized as a fundamental human right and as foundational to the exercise of other human rights**

**What challenges, if any, would be created for organizations if the law were amended to more clearly require that any development of AI systems must first be checked against privacy, human rights and the basic tenets of constitutional democracy?**

<u>**Short reply**</u>**:**

A list of potential challenges can be found below:

- Increase in internal infrastructure, organization, and compliance costs
- Increase in HR and IT costs
- Potential uncertainty and ambiguity as to interpreting and implementing "appropriate" and "reasonable" measures and safeguards
- Chilling effects for investment
- Compliance-based access barriers to market
- Different scales of sophistication and equity or reserved funds (micro-enterprise, small, multinational bank, etc.)

<u>**List of Recommendations:**</u>

1. **Consider augmenting policymaking by incorporating structure and semantics of coding language for software systems when imposing obligations for appropriate and reasonable measures and safeguards. In other words, semantics of coding and semantics of human language could be better aligned to mitigate loss of translation risk when such legal rules are coded within a system, for privacy or compliance by design or by default purposes. This proposes shared vernacular between the legal and technical domains.**

2. **Consider public sentiment and expectations to determine clear Go and No Go zones for AI systems intended purposes as well as for data protection**

Montreal AI Ethics Institute    13

3. **Consider augmenting policymaking by designating a multidisciplinary group of policymakers, including technical and business staff involved in the design, development, deployment and maintenance of AI systems, lawyers and ethicists.**

4. **Consider building and promoting data protection and systems literacy through regulators' channels and websites, for education, coaching and training purposes. Such citizen friendly bodies whose main or explicit purpose would be education and public competence-building would improve trust[11].**

5. **Consider innovation and compliance sandbox environment to enable or facilitate transition across access barriers.**

**Comments:**

Human rights have been constitutionally advocated as foundational to any and all activities happening within a given jurisdiction -in this case, Canada-, but in practice, it mostly appears the analysis and risk management models implemented for monitoring and assessing the respect of human rights have often been siloed into the human resources **("HR")** divisions of businesses and corporations, if any, and not as foundational to corporation-wide strategy as a whole.

We consider explicitly prioritizing a human-rights based approach to data protection law in Canada has become critical. We are aware of potential challenges which may find friction with businesses in implementing such an approach, and they have been identified below.

---

[11] See Montreal AI Ethics Institute, https://montrealethics.ai

Montreal AI Ethics Institute    14

**Additional compliance, HR and organizational and IT infrastructure costs for businesses:**

During the course of the past few years, we have slowly seen different corporate positions and governance structures emerge or being debated, as the disruptive nature of algorithmic systems, which may output biased results and thus exacerbate illegal discrimination issues, has made ethical and human rights considerations resurface as a priority -or, as an important business or market risk-. Such positions have titles such as Ethics and Compliance manager, and we have seen either new internal structures which have designated ethics and human rights committees or independent or semi-independent advisory boards, to assess such risks, or to meet transparency and compliance obligations.

As such, where traditionally human rights issues were dealt by HR for specific employment issues, human rights assessments are slowly spilling out of the HR division and throughout the corporation consequently to the implementation of systems and algorithms which require a more thorough and corporation-wide assessment of human rights risks.

Evidently, this "spillover" comes with additional compliance and HR costs, through the implementation of new budget allocations, for new teams, structures, protocols and processes, and additional organizational and technical measures. Time and efforts are spent on such governance transformation instead of market share increase, augmenting market penetration, increasing sustainable growth, or profits' generating activities. Research for existing expertise in the matter as well take a lot of time, and although such expertise can be developed internally through coaching and courses -which comes with its own set of costs-, perhaps an external expert would be advisable, but depending on availability, it may also increase HR costs if demand outweighs supply.

It may be argued that strengthening the Canadian data protection legal framework by increasing compliance costs may have a chilling effect on investors and venture funds, which may directly impact the Canadian economy, pace of innovation and put up access barriers to the software technology industry. However, it must be kept in mind that if such a framework may be expected and demanded by Canadian individuals and aligned to their particular concerns and ethical positions, then such a framework merely reflects the rules of the game which must be respected by businesses to provide products and services within the country.

Such businesses would otherwise in the short or medium-term face rising conflicts and tensions with citizens, clients and employees, if the products and services and automated systems they deploy upon them, which would inevitably increase customer service costs, costs to deal with increased churn, loss of clientele, reputation damage, or litigation claims. We have seen many of such boycotts and citizens groups protests in the United States and other countries, where specific projects went against communities' beliefs and ethics[12].

**Uncertainty of degree or level of appropriateness or reasonableness of legal interpretation and implementation of human rights organizational and technical measures outside of HR context:**

This is not a new challenge, but it is an ongoing one, which may have consequences per legal provisions, litigation, repetitional damage, and compliance costs, including that which is stated in public and private financial statements.

When it comes to human rights legislation, there is case precedent and historical directives and regulatory guidance for interpretation and applicability. However, charters and other human-rights legislation are often general in nature, and provide from broad, "elastic", flexible scopes, to ensure that they may evolve as "a living tree"[13], and remain efficient with, and representative of, evolving times. Historical jurisprudential precedent may assist in adequately framing such corporate governance, but there is a lack of interpretative guidance for AI systems which may hinder corporate and regulators' data protection audits and impact assessments.

There are pros and cons with such elasticity, the main advantage being that individual cases should not normally fall beyond jurisdiction or would be automatically excluded, which provides fairness and democracy within the system. The disadvantage, however, is that in terms of implementing appropriate technical and organization measures and safeguards, it may be difficult to interpret and meet the "appropriateness" threshold. The issue also applies in terms of identifying what would be "reasonable" measures or safeguards, depending on the level of sophistication of the business. A micro-enterprise does not have the equity or reserved funds for compliance structures in the same manner as a traditional multinational bank, for instance.

As such, we are aware of the difficulty for regulators and legislators alike to determine in advance what would be appropriate or reasonable, in terms of identifying specific

---

[12]CRAWFORD, Kate, others, *AI NOW 2019 Report*, December 2019, 100 pages.
[13]*Reference re Same-Sex Marriage*, [2004] 3 S.C.R. 698, 2004 SCC 79, "living tree"

organizational or technical measures and safeguards, to provide data privacy protection, human rights due diligence and protection, and cybersecurity. The issue is exacerbated in circumstances where micro-enterprises may quickly impact individuals on a systemic level in a matter of days. An analogy can be driven from fintech's developments, where some start-ups can become "too-big-to-fail" entities in an extremely short amount of time[14]. Furthermore, technological developments are apace, and as such, imposing specific techniques in a regulation may quickly become obsolete.

However, it appears from our experience that public sentiment wishes and demands that concrete positions soon be taken by legislator and regulators alike. This is why understanding the level of comfort, security, concerns and acceptability of AI within specific local cultures and communities, before drafting a legal framework has become key. Considerations for strict and clear "Go and No Go zones", based on local public endorsements or ethical declarations, may be important, to better direct businesses for future developments.

Finally, there needs to be an emergence of a new class of discipline and employees who are equally well versed in the technical and legal aspects because apart from that, no regulation can be effectively implemented. GDPR demonstrated the difficulty in amending systems from a technical perspective and paying a greater degree of attention from an AI perspective at time of policymaking means we need to consider how practical it is to be able to adhere to and address the advances in the technical domain and square those with the law which might not itself be fully future proof given the pace of change in the field.

---

[14]ARNER, Douglas, OTHERS, "Fintech and Regtech in a nutshell, and the Future in a Sandbox", CFA Institute Research Foundation,
https://www.cfainstitute.org/-/media/documents/article/rf-brief/rfbr-v3-n4-1.ashx, p.7.

**Proposal 3: Create a right in the law to object to automated decision-making and not to be subject to decisions based solely on automated processing, subject to certain exceptions**

1. **Should PIPEDA include a right to object as framed in this proposal?**

**Short reply:**

We suggest the intent in drafting a right to object be reoriented into the intent of explicitly referring to a **"right to negotiate with a human"**.

**Recommendation**:

- **Provide for a right, or mechanism, of negotiation with a human for an individual concerned once a decision is made by an AI system about such an individual. This negotiation mechanism is proposed as a positively drafted amalgamation of the right to object and the right of appeal, as well as parts of a right for reasonable inference, but better aligned with practical business relations and interactions. See proposed design and test below under #2.**

**Comments:**

**General technical challenges linked to the right to object:**

In the right to object to automated decision-making there are a number of concerns. First, explicit consent is hard to be informed when there is a complex system that uses many factors to make a decision.

Second, there is nothing like sole automation, because even when automated systems are used to render decision support, there may be temptation in avoiding restriction for sole automation by appointing a "human in the loop" as a "token human"[15]. Additionally, such human analysts, even if not a token at the beginning of such mandate, may become prone to overreliance and inertia biases, practically deferring all cognitive logic and decision-making to the systems, unless adequate measures are not imposed. In essence, such processes may become automated decisions in the first place, and subtly can subvert the law that relies on the sole automation clause.

---
[15]See GDPR, and EU Guidelines for Trustworthy AI,
https://ec.europa.eu/digital-single-market/en/news/ethics-guidelines-trustworthy-ai

Furthermore, contesting the decision requires an understanding on part of the data subject in terms of if they might have received a fairer prediction, alluding to explainability of the system which is a hard technical problem.

There is also a risk of an automated system being submerged under other benign non automated systems. For example, an upstream API using AI could hide a certain subset of nature of the system being automated.

Human intervention itself can only be meaningful when a person understands how the system operates, what resources they have to pry open the system to understand it, access and transparency on what data was used in the training of the system and what residues that has in the context of the current decision being made.

In addition, once data is incorporated into the training, there is a tremendous amount of difficulty in extricating the impact that has on the trained model which might make it infeasible though there is some progress on that front[16].

Correcting information about themselves in a machine learning system is extremely hard because of how much impact an individual's data has on the trained model is not easily quantifiable and its extrication is a very challenging problem.

Similarly, requesting information that is not held in a dataset anymore but in a trained model suffers from the same problem.

**Limitations of the right to object, and Negotiation as a key social and contractual human interaction:**

In the manner the present proposal is drafted per the right to object, there appears to be an assumption that human-based decisions will be more just, more fair, or more favourable to another human than a machine-based decision. Similarly, the drafting of GDPR appears to provide for a general prohibition of automated decision-making systems, except for such authorized processing activities with stringent requirements.

---

[16]BOURTOULE, LUCAS, OTHERS,, Machine Unlearning, University of Toronto and Vector Institute, https://arxiv.org/pdf/1912.03817.pdf., 16 pages.

It is certain that a human-based decision appears more flexible in a preliminary manner, than fixed parameters or variables embedded within an automated system. In the same manner, it is certain negotiation with a machine that has been programmed to provide for specific scoring or results with strict or evolving criteria or parameters will be difficult, if not impossible, whereas the apparent opportunity to at least attempt or be able to negotiate a price or decision with another human provides a certain sense of security, sense of power, control, or reasonableness to the consumer. One must note however that humans may be inflexible as well through their decision-making, whether through prior prejudice, bias and personal beliefs. In the ignorance as to how the machine has gone through an individual's information to identify the decision for him, the individual has at first very little visibility or understanding as to its inner workings.

For example, in South Korea, students and job applicants are now taking extracurricular classes to learn how to "best" or "beat" an AI hiring bot equipped with NLP and sentiment analysis during a preliminary interview. They are thought to master specific expressions to "fool" the bot and attempt to be provided with a higher score or pass the preliminary interview to obtain a human second interview[17]. A problematic response to this deployment is for instance the pressure put on individuals to try and identify how a bot or AI functions when they are being screened.

This is not particularly different from negotiating with the blackbox of another human's brain, but between an imperfect NLP algorithm with approximate sentiment analysis understanding, which takes away from the interviewee any opportunity of "reading" his robotic interviewer's reactions to adequately nuance or orient his proposal, and a human with inherent biological and cultural recognition of such, trust and credibility remains higher for a H2H (Human-to-Human) decision than a H2M (Human-to-Machine) one.

It is also certain that the "elasticity" of both the collection of data and negotiation of a business relationship may be greatly reduced by automating these with software and machines, unless prospective customer has knowledge and access to specific NLP or software /platform loopholes, or to critical or weighed data labels which could redirect machine decision in its favour. Of course, the flipside of too much transparency as to the features of a specific system is the augmented risk of adversarial attacks against the system, which can game the process and create unfair advantage that will not be far from how people might curry favors via bribes, although, strictly speaking, the latter are more explicitly unethical than attacks against a system. A prime example would be a situation called "zero day" attacks in cybersecurity, where a hacker may discover a vulnerability or weakness but does not otherwise trigger immediate exploitation. They can and often do exploit the system for gain but do not disclose the vulnerability to the people behind the affected system. As there are today, systems may lack robustness, and may remain weak and brittle against such adversarial attacks.

---

[17] CHA, Sangmi, "*South Korean jobseekers and students are beating the AI interview bots - here's how*", World Economic Forum, 2020,
https://www.weforum.org/agenda/2020/02/south-koreas-ai-hiring-bots/

In practice, the spectrum of this human "negotiation opportunity" does vary depending on the business model and business culture. For many sophisticated or big corporations, a certain level of standardization of protocols and decision charts is already provided to first line of defence, customer service, or sales representatives, as they meet with prospect or recurring customers. Despite perhaps a certain appearance of human negotiation, at times these representatives have little margins for negotiations as they are imposed specific targets by their employer, and exceptional discounts or deals then require explicit authorization from management or director. In other businesses or cultures, negotiation is an intrinsic part of everyday business, and everything can be negotiated. In an ideal market world, both extremes, and everything in between, should be left as permissible.

It is important to keep in mind that classifying prospect of recurring customers into specific profiles or categories of persons based on data or statistical methods is not a novel business practice. Already, much marketing analytics is done and made available to business lines to create such profiles based on key characteristics or inferences drawn out of pattern-like or recurring behaviours.

Adding automated decision-making systems will not change the fact that such key characteristics and rules to output business results or decisions, based on such business inferences, whether deemed reasonable or not, may already have been decided by marketing statisticians, behavioural economists and business strategy teams prior to entering them within the software as determinant variables or parameters.

What is new is attempting to obtain in real-time more personalized profiles for each individual concerned by integrating available behaviour data from different data ecosystems (ie. social networks, for instance) into a system, and providing more flexibility to the systems through machine learning, for it to continuously calibrate or update the individual's profile with newly available data. This may mean permitting the systems to add new parameters or factors to its logic, or modifying weights provided to different existing parameters.

Academics have looked into the right to reasonable inference[18] instead of, or as a corollary to, the right to object. In this case, an individual would be provided with visibility and a certain level of control in the logic programmed within the systems, to determine whether or not the result is reasonable and fair because of a fair and reasonable logic or inference.

We consider this right will not resolve the problem, since one will be confronted with the difficulty of identifying what is considered reasonable or not, and which may appear first as a subjective test. Plus, it is also important to maintain the market freedom to let corporations make business inferences based on different categories of data.

For instance, although arguments could be made that a reasonable business inference would be most aligned with the market needs and does not necessarily mean a reduction in profits, business inferences, whether made by a human or a system, ought to be permitted to be reasonable or not, but may not be based on such practices impeding rights and freedoms of individuals, or equal to abuse under such common law or civil claims available (ie. consumer protection). There is a certain "elasticity" and a certain "arbitrariness" in business which must be able to coexist and be respected, irrespective of the fact they may appear reasonable to one, and unreasonable to another.

It is also important to keep in mind that discrimination between different individuals may be legal, unless it has been explicitly identified as illegal per charters of human rights and freedoms. For example, if a bank provides for a business inference that scores an individual's mortgage loan interest rate based on the fact that he comes from a different ethnicity, or based on the behaviour of other members of the same ethnicity, and another individual from a different ethnicity would either be favoured or disfavoured based either solely on this factor, or whether such factor was weighed into the decision logic, then this business inference is evidently illegal and as such, the systems ought not have been designed in such way to permit for such risk to arise, and, in the event the complete avoidance of such risk is not possible, to at least create escalation protocols and compensation mechanisms for individuals which may be adversely affected by such decisions.

---

[18] WACHTER, Sarah, MITTELSTADT, Brent, "*A Right to Reasonable Inferences: Re-thinking Data Protection Law in the Age of Big Data and AI*", Columbia Business Law Review, Vo. 2019, p. 495-620, 127 pages.

But for example, if a bank decides to augment the traditional data provided by credit scoring agencies by adding behaviour data of the subject on social networks because research would hypothetically show that the credit risk profile would be more accurate, then in this respect, even if some inferences may seem unreasonable or incorrect to the individual, the bank should have the freedom to develop its own systems' logic. Of course, arguments can be made that consent of the individual needs to be provided for the social networks behavioural data, in the same way it was explicitly provided for the credit scoring data, and of course, caution is advised since proxies of data can be encoded into data from social networks.

The right to object is drafted in a negative manner. An individual may be unhappy with a business decision taken about him or her, but it does not make this business decision or business inference illegal, unless once again it impedes existing rights, freedoms, or meets the requirements for a consumer abuse claim or other claims at law. This is why we suggest instead the right to negotiate with a human, because with the right to negotiate, the right to sufficient or meaningful explanation becomes its corollary, as sufficient information is required for an individual to be able to negotiate in a meaningful manner. We identify negotiation as an intrinsic and critical human behaviour. Whether consciously or not, an individual is continuously negotiating sensory data, and weights and priorities applied to these and to his self narrative script, between his internal "map" and the external world, the importance of his self determination against collective good, expectations and social and cultural norms, as well as everyday when interacting with other individuals, whether in a personal or business context. As such, this interaction should be protected at law, since negotiating with a system is not "humanly" possible. And in the context where a business negotiation fails, the individual concerned may rely on other civil claims and recourse available, if there is sound basis that the decision-making was done illegally.

In this respect, under the right to negotiate with a human, the individual concerned will be provided with the reasons and general factors underlying the business decision, automated or not, and can attempt to justify how certain factors should be changed, if he considers these as erroneous or unreasonable, or he may be provided the opportunity to provide new or other data to assist in the business decision to go in his favour. However, it is to note that the right to negotiate does not otherwise mean that the individual must always obtain a more favourable decision with a human.

2. **If so, what should be the relevant parameters and conditions for its application?**

**Reply:**

**An example as to how the right to negotiate with a human would be processed is proposed as follows:**

1. An individual A wishes to obtain a product or service
2. There is an automated decision-making application -identified as such- on the website of a business providing such product and services
3. Mandatory prior disclosure details per functioning of the system and manner in which data will be handled and processed are provided for A's meaningful consent.
4. A consents and opens/triggers AI application
5. A provides data strictly necessary for AI application to output result
6. a) Result is to the satisfaction of A ——- A proceeds with next steps.

   b) Result is not to the satisfaction of A
       i) AI application may provide for reasons and key factors for the result, as well as contact information details and protocol for negotiation and verification with a business representative. AI application may also provide for what additional steps or information could be provided to obtain a better, or alternative result.
       ii) Business representatives must be sufficiently adept at reading AI application findings and should be equipped with controlled access to decision-making, and other negotiation tools (it cannot be a human token).
       iii) The minimal negotiation test obtains a successful score or result (see below)

7. **Optional:** AI application asks A whether it can use other databases, profiles or accounts to further personalize result.
8. Provide additional mandatory prior disclosure, as described under #3 above.
9. Obtain consent as #4 above. (the consent is specific to the use of the AI application for that specific decision only)
10. a) Result is to the satisfaction of A ——- A proceeds with next steps.

    b) Result is not to the satisfaction of A: proceed as under #6b).

**The minimal negotiation test proposed for such right to negotiate would be three-fold:**

1.-At the time the automated decision is made, can the business representative verify and guarantee that the decision in question made by the system is not founded on illegal discrimination, privacy, or human rights and freedoms?

2-What were the determining factors for the result provided to the individual, and would they appear both objectively and subjectively reasonable?

3-Whenever possible, what would be the additional information or subsequent actions the individual could take to modify the result and enter a contract with the business?

**Note 1:** Specific organizational and technical measures should be put in place to ensure the business representative will not be, and will not become over time, negligent or over reliant upon the findings of the system without providing a real chance for meaningful negotiation with the individual (mitigate risk of inertia and over reliance biases, as well as augment atrophy of cognitive skills, amongst others).

**Note 2:** Consideration should be provided for such standardized contracts at law which are inherently non-negotiable. Explicit mention that the negotiability of the test above does not otherwise change or modify the nature of a non-negotiable or unilateral contract, would be recommended.

**Note 3**: The right to negotiation comes also as an alternative mechanism to fix or resolve the intractable problem brought about explaining complex systems, associated with the right to explainability. In other words, regardless of the result, an individual has the means to have his personal request considered anew, regardless of his understanding or lack thereof, or that of the explainability or lack or complexity thereof, of the automated system.

**Proposal 4: Provide individuals with a right to explanation and increased transparency when they interact with, or are subject to, automated processing**

1. **What should the right to an explanation entail?**

**Reply:**

In the legislation concerned, there may be some information disclosed to the end-user, as well as a right to explanation concerning the logic used by an AI system, but for complex systems that is an intractable problem. As such, we propose that, standing alone, such right -or mandatory minimal disclosure obligation imposed on businesses- must be supported by other mechanisms, such as the right to negotiation, as described above.

Whether it be called a right to an explanation or a mandatory minimum disclosure from businesses concerned, we propose elements to be disclosed as follows:

**Disclosure of:**

- The use of an automated system by a business (whether it be to take business decisions towards the individual, an NLP chatbot, or otherwise), if it may have a direct or indirect effect on the individual, irrespective of whether this effect may be positive or negative
- The intended purpose, as well as a use-based privacy approach[19]
- The proper method of using it (if individual is able to do some manipulations of the tool through an interface)
- The general method of how it works (weights of key factors, and if the machine learns, the general changes that can be expected)
- Whether or not the individual has access to controls or settings to reset the ML system pertaining to his profile or account, and where to find them (should be easy)
- How the individual can restrict or withdraw consent for the types, sources, and categories of data used by the system to make a decision, and the pros and cons of doing so -unless information is required by law, individual should not be forced to either accept or refuse the business product or service on the basis that he must provide more data than that which is immediately required for the task.

---

[19] BAGDASARYAN, OTHERS, "Ancile: Enhancing Privacy for Ubiquitous Computing with Used-Based Privacy", Cornell University, 2019, http://www.cs.cornell.edu/~jnfoster/papers/ancile.pdf

- A comparative approach could be tested, as in provide for what would the output results are or would be, with or without additional data, and provide the individual with the option then to choose which is most favourable.
- The types and sources of data that have been used to output the results
- **How the privacy of the data is being guaranteed or managed (make it insufficient to simply provide a link towards the privacy policy of the business, the information need be particularized per the specifics of the system being used and affecting the individual)**
- The level of risk for illegal discrimination or bias within the system based on the data used, and the actions the individual can from then on take to modify such decisions
- What are the expected results supposed to be (eg. Yes, No, additional information required)
- Either provide the actions and information missing if the individual is not given the result he was looking for automatically through the software, or through a human channel.
- The consequences of such results for the individual
- The manner in which the individual may be provided with additional details about his own personal profile, account, or manner in which his data has been used by the automated system
- The manner in which the individual may meaningfully negotiate the output results with a human if he disagrees with it or thinks he should get a better deal, and also as a means to verify how the system dealt with his data and profile (meaningfully means it is not sufficient to merely place a token human, individual must have the actual freedom of being able to change the output result towards him through negotiation if he meets other criteria)
- **This should ideally be done prior to the use of the system, and with explicit consent from individual (eg. checkbox)**
- Should be provided in clear, concise and in simple language (ie. keep it short and simple principle)
- A multi-channel approach in terms of decision-making would be optimal, to ensure a better transition of individual to the automated systems, ie. there should be no illicit business pressure reorienting individuals towards the automated systems without any other alternative avenue of providing the data and being provided with a decision. (Multi-channel).

**Note 1:** Careful consideration should be provided to public feedback provided during consultations, for such elements or factors of disclosure which a given target audience or local culture, community or demographics wish to be provided with. This permits to add more fine grained definitions for individuals, and the reasoning required from the system can be tailored to the level of understanding of the target audience. As such, to complement the proposal of different factors above[20].

2. **Would enhanced transparency measures significantly improve privacy protection, or would more traditional measures suffice, such as audits and other enforcement actions of regulators?**

**Short Reply**:

A combined approach is best. Yes, enhanced transparency measures would improve privacy protection, but they must be accompanied with more credible, deterring and efficient traditional and regulatory measures pertaining to audits and enforcement actions. Furthermore, consideration should be provided to prescribe certain technical fairness and accountability metrics to support the transparency measures.

**Comments**:

Enhanced transparency measures would provide added incentives to business players to better audit, monitor and verify ongoing business partnerships and provide for more enforceable representations and warranties from their business ecosystem due to the added risk to potential reputation damage if one of their key business players are not compliant with legislation or their own level of protection. This needs to be linked to the explicit assurance that all business players which collect, receive or otherwise process or transfer personal data remain liable towards the individual concerned in case of violation or breach of privacy for full damages and other measures. The "chain" of liability must remain solid throughout the ecosystem, as it is the only incentive ensuring business players will work with other responsible business players.

Pertaining to AI systems, transparency, fairness and accountability all require a certain degree of technical specificity for them to be useful. In other words, mere principles explicitly identified at law will provide behavioural guidelines, but will most probably be inefficient if certain metrics are not prescribed to be embedded within the AI systems.

---

[20]See public feedback gathered by MAIEI in Schedule 1; ZHOU, DHANKS, "Different Intelligibility for Different Folks", https://www.aies-conference.com/2020/wp-content/papers/088.pdf

Relative to fairness, most current work in machine learning focuses on application of restricted definitions on static datasets assuming that the distribution of data remains unchanged throughout the utilization of the system which is not the case for complex, real world systems that evolve in their interaction with new, incoming data.

While this sort of analysis may work for systems in simple environments, there are cases (e.g., systems with active data collection or significant feedback loops) where the context in which the algorithm operates is critical for understanding its impact. In these cases, the fairness of algorithmic decisions ideally would be analyzed with greater consideration for the environmental and temporal context than error metric-based techniques allow. Actions informed by the output of the ML system can have effects that may influence their future input.

As an example, the thresholds determined by the ML system are used to extend loans. Whether people default or repay these loans then affects their future credit score, which then feed back into the ML system[21]. There is an inherent tradeoff in the fairness definitions as it applies to subgroups and true positive rates among other metrics[22].

Accountability is probably the one aspect that requires the least technical specification yet needs to be fundamentally incorporated into the development and deployment processes of AI systems.

The transparency requirement is central to the enforcement efficacy of any of the measures including that of audits and regulators. Even if full transparency is not required, audits require that to some degree for them to be meaningful.

However, it is absolutely vital that relevant authorities concerned in the audit, supervision and enforcement of the privacy legislation be provided with at least the basic minimum enforcement powers as recognized by law to ensure that there is sufficient credibility as well as capacity in applying the law.

---

[21]SRINISAVAN, Hansa, ML-fairness-gym: A Tool for Exploring Long-Term Impacts of Machine Learning Systems, Google AI Blog, , http://ai.googleblog.com/2020/02/ml-fairness-gylem-tool-for-exploring-long.html

[22]KLEINBERG, OTHERS, "Inherent Trade-Offs in the Fair Determination of Risk Scores", https://arxiv.org/abs/1609.05807; AI Fairness 360 Toolkit, IBM, https://www.ibm.com/blogs/research/2018/09/ai-fairness-360/

**Proposal 5: Require the application of Privacy by Design and Human Rights by Design in all phases of processing, including data collection**

1. **Should Privacy by Design be a legal requirement under PIPEDA?**

**Short Reply**:
Yes. If Privacy by Design is not imposed as an obligation at time of design and throughout the lifecycle of the product and in the normal course of their activities, businesses may not feel incentivized in making this mindset a priority. Furthermore, consideration should be given to impose a Human Rights by Design legal requirement as well.

**Comments**:

General principles pertaining to Privacy by Design are normally associated with principles or rules as follows[23]:

- The privacy assessment and measures need be proactive and preventive, not reactive or remedial
- Privacy measures are imposed as both technical and organizational from the start or at time of design of the product or system
- Such protective measures must provide for "end-to-end security"
- There is a requirement for sufficient visibility and transparency
- The respect of user privacy is foundational to all technical or business projects

As for the human rights by design approach, the UN Guiding Principles for Business and Human Rights is a preliminary resource which can be useful for both states and businesses in designing the appropriate public and corporate governance framework in such respect, from time of design to post-mortem review of governance structure and evolving human rights case scenarios relevant to the business.

---

[23]CAVOUKIAN, "Privacy by Design: Foundational Principles",
https://www.ipc.on.ca/wp-content/uploads/Resources/7foundationalprinciples.pdf; Search Encrypt, Medium, "7 Principles of Privacy by Design",
https://medium.com/searchencrypt/7-principles-of-privacy-by-design-8a0f16d1f9ce

As an example, the UN indicates the following, amongst other[24]:

- Corporate statement and commitment for responsibility in ensuring the respect of human rights
- Implementing appropriate prevention, mitigation and remediation measures to address adverse human rights impacts
- Commitment to support and promote human rights is commendable and should be encouraged, but it is insufficient to ensure appropriate operational measures business-wide
- In defining human rights concepts, various useful sources are the Canadian and Quebec Charters in such respect, as well as the International Bill of Human Rights
- Obligation for periodic review of corporate human rights governance
- Implement policies and processes factoring in business size and circumstances, as well as potential severity of impact to human rights, including a human rights due diligence process
- Public transparent disclosure of human rights corporate governance policy and processes, as approved by senior level of business concerned
- Other.

However, additional guidance and standards pertaining to operational and technical measures should be explored and proposed based on Canadian legal and economic ecosystems' particularities.

Considering the above, we are of the opinion that unless such privacy and human rights by design are not made a mandatory requirement in all business cases -but explicitly referred to AI systems- to embed available and existing techniques within systems, plus imposing certain organizational measures throughout the engineering, manufacturing, or scaling process, from beginning (design) to end, business players may chip and nibble at the required governance threshold to attempt to save costs and take more risks which may be detrimental to individuals concerned. Imposing the framework/techniques of privacy by design or privacy by default, or an essentially similar framework/techniques, throughout the chain, is a most optimal way to provide a certain sense of safety and relative guarantee to individuals that their personal data goes through systems built to keep the data private.

---

[24] ALLISON-HOPE, Dunston, BSR, "Human Rights by Design", https://www.bsr.org/en/our-insights/blog-view/human-rights-by-design; PENNEY, Jonathon, OTHERS, "Human Rights by Design", Schneier on Security, https://www.schneier.com/blog/archives/2018/12/human_rights_by.html; Office of the High Commissioner for Human Rights, United Nations, UN Guiding Principles for Businesses and Human Rights, https://www.ohchr.org/Documents/Publications/GuidingPrinciplesBusinessHR_EN.pdf

There are concerns as to which appropriate definition, general principles and foundational rules need be identified within a privacy and human rights by design obligations imposed on businesses, and the scope of such rules. Furthermore, a requirement for certification mechanisms is often associated with such designs, and clear standards will need to be identified. Finally, tests and oversight assessments will need to be identified if they are to be implemented with a privacy by design process. Further exploration in terms of defining prescribed technical and organizational requirements is required, in particular in terms of balancing pros and cons associated between strict standardization of process, and ensuring sufficient flexibility to encompass present and potential case scenarios.

2. **Would it be feasible or desirable to create an obligation for manufacturers to test AI products and procedures for privacy and human rights impacts as a precondition of access to the market?**

**Short Reply**:
Yes it is feasible, and yes it is desirable, both as technical and organizational measures.

**Comments**:
Guidance should be provided as to the manner and method of enabling such testing. Of course, we would recommend explaining the difference between the obvious need to test the machine-learning model after training, versus the testing as prototype in a controlled and less controlled environment on, chronologically:

**Design:**

1. Testing the intended design and purpose of the model against existing illegal bias, fairness, ethics and human rights, and cybersecurity safeguards due diligence procedures
2. Testing the intended design and purpose of the model within a risk management assessment framework (ie. Data Protection Impact Assessment **(DPIA)**)

**Post prototype Build:**

3. Test as under Design once more to ensure there have been no changes

**Testing of prototype model:**

4. Controlled sandbox testing on volunteer employees, then tasked to provide feedback and comments + monitored and assessed by programmers
5. Controlled sandbox testing on volunteer clients with particular incentives, which are tasked to provide feedback and comments + monitored and assessed by programmers.

*Ongoing monitoring governance protocol should also be implemented.

Linked to the right of explanation or of additional disclosure pertaining to systems, the explicit confirmation provided to individuals concerned by manufacturer/business players that the system has undergone and passed the testing protocol.

From a technical perspective, an open question that comes to mind is how to prove compliance with the law especially of systems that are developed under different legal systems and that might have those legal notions baked in and can't be extricated or amended in a meaningful manner thus preventing their use. This might be particularly problematic in the use of tools that obviously lend to the population using it benefits that have come to be expected as natural in the context of our digital existence.

Furthermore, providing for a rule stating businesses concerned must avoid any potential bias may not be identified as meaningful language because in some contexts having bias in them makes sense, for example, when there is a higher incidence rate of heart attacks in men vs. women. If we do not account for the difference in distribution of this in the system via an equitable dataset, we may end up with a system that underperforms on demographics because of a false notion of all bias prevention. Biases should be thought of in the context of the application in consultation with a domain expert and then addressed meaningfully such that outcomes are properly accounted for and not just relying on the fact that biases were removed in the first place that there might not be proxies that can negatively harm the outcomes.

Finally, the test authorizing access to markets should also extend to such products that use upstream APIs that can subtly let slide in biases and other privacy issues, and not only AI systems. As such, procurement processes for these products need to have that as a requirement as well because some systems might originate in an extra judicial setting where they are not subjected to the same regulations, for example, when using face recognition APIs based on the models developed in some other jurisdiction.

**Proposal 6: Make compliance with purpose specification and data minimization principles in the AI context both realistic and effective**

1. **Can the legal principles of purpose specification and data minimization work in an AI context and be designed for at the outset?**

**Short Reply**:

Per purpose specification, the answer remains yes, although it is an insufficient measure and requires additional mechanisms or measures in such cases where limiting a business application or research project cannot be practically secured. It should be expanded to include disclosure of intended and concrete purposes of an AI system.

Per data minimization, it does not adequately consider inferences made from input or collected data, especially relative to aggregated trends or the mosaic effect[25]. This may harm the efficacy of this principle. It should be supported by use-based privacy and systems thinking approaches[26].

**Comments**:

It is crucial to have the obligation to limit and disclose data purposes remains to limit the design, use and scaling of AI systems to such purposes that have been fully and meaningfully disclosed to the individual concerned, and has been provided clear and enlightened consent for the collection, use, processing, conversion or transfer of the personal data being inputted and outputted by the systems.

It is argued that at the time of design of an AI system, all relevant potential purposes cannot be identified and as such the purpose limitation fails. In our view, the exercise of identifying and selecting the intended purposes of the AI at time of design, to be complied with and respected through the remainder of the engineering and marketing process, is a very good exercise. The legal mechanism where a new intended purpose is considered, then brought to the attention of the individual for consent purposes, must remain. It is common sense and expected from an individual for its personal data not be used for other means, or in other ways or purposes without his or her consent, in particular if it may have an effect on him.

---

[25] HUBER, Rose, "Mosaic Effect" Paints Vivid Pictures of Tech Users' Lives",
https://wws.princeton.edu/news-and-events/news/item/mosaic-effect-paints-vivid-pictures-tech-users-lives-felten-tells-privacy
[26] MEADOWS, Donella, "Thinking in Systems",
https://books.google.co.in/books/about/Thinking_in_Systems.html?id=CpbLAgAAQBAJ&redir_esc=y

Per data minimization, it is slightly more difficult in practice. ML systems require constantly updated new data to improve. Machines do not learn in the same manner as humans. But these methods and techniques per which data can be relatively de-sensitized or converted into neutral data, which could then be fed to the AI systems. Some will say that cleaning the data in a manner that would de-personalize creates a detrimental trade-off, but it should be possible to minimize the personal data to what is strictly necessary for business and services rendered, and then provide the option to the individual for added personalization per his consent. Not all AI systems or personal data create necessary plus-value for businesses or society. Same with personalization of output results, it can be difficult to quantify the exact added qualitative and quantitative value of this mechanism.

Something that is not explicit in the data minimization principle is the role of data proxies which can evade the purpose specification clause and lead to problems. In addition, data that is not sourced from the individual and thus avoids the judicial weight of these two principles, for example, utilization of datasets collated by data brokers[27] that largely operate outside of the legal boundaries, do not appear to be clearly accounted for, either in the definition of personal information, or within applicable principles or rules of PIPEDA to date. A similar comment can be made for such "new" data created or inferred by AI systems based on individuals' data. Some arguments have been made that data minimization may be excluded from PIPEDA principles through the use of de-identification techniques. However, there are so many examples[28] where de-identification of personal data is grossly insufficient as a means of privacy protection.

Per the argument that there would be reduced quality from limiting personal data collection, it is not necessarily the case. Specifically, transfer learning can allow for training on non-sensitive or public data.

---

[27]FEDERAL TRADE COMMISSION, "Data Brokers: a Call for Transparency and Accountabliity", 2014, https://www.ftc.gov/system/files/documents/reports/data-brokers-call-transparency-accountability-report-federal-trade-commission-may-2014/140527databrokerreport.pdf
https://www.ftc.gov/system/files/documents/reports/data-brokers-call-transparency-accountability-report-federal-trade-commission-may-2014/140527databrokerreport.pdf
[28]GUPTA, Abhishek, "The Evolution of Fraud: Ethical Implications in the Age of Large-scale Data Breaches and Widespread Artificial Intelligence Solutions Deployment", https://www.itu.int/en/journal/001/Documents/itu2018-12.pdf

For example, it is possible to train and test a model on data about individuals that are no longer alive and who will not face any harm from use of such data. This requires legal expiration policies -which already exist within PIPEDA- beyond which data joins the public domain, subject to specific types of data, such as genetic data, or potentially hereditary health data, which may retain relevance for individuals other than those who are no longer alive.

An example of this can include training on ImageNet[29] to glean basic characteristics of a large category of objects. Then this pre-trained model can be used on other downstream tasks by fine tuning the model.
Another example is large scale language models like GPT-2[30] that capture basic language semantics and grammar and can then be fine tuned for domain specific tasks.

Finally, both principles could be augmented by integrating a systems thinking approach, whereby relevant stakeholders and externalities can be identified, and this may enrich the analysis and the degree of efficacy.

2. **If yes, would doing so limit potential societal benefits to be gained from use of AI?**

**Short Reply:**

In our view, at this point in time, the importance of reassuring the public that systems will be designed responsibly and will not intrusively use personal data to control or nudge them into specific behaviours or locking them in specific profiles, and to mitigate risk and fears associate to exacerbating social biases and illegal discrimination, far outweighs the hypothetical loss to society associated with eliminating the purpose limitation and data minimization obligations.

3. **If no, what are the alternatives or safeguards to consider?**

Please see comments under Proposal 6, questions 1 and 2.

We suggest augmenting existing principles with use-based privacy and systems thinking approaches. In this respect, it will not limit the amount of inferences made by AI systems, but how such inferences can be used.

---

[29]IMAGENET, http://www.image-net.org/
[30]GPT-2, https://openai.com/blog/gpt-2-1-5b-release/

Montreal AI Ethics Institute                                36

In addition, we would like to clarify that the data minimization would be more useful when applied to collection of data, but not necessarily to limit the "new data" built through the machine-learning system and pipeline, as inferred from input data. For this "new data", purpose limitation and use-based privacy would be optimal.

**Proposal 7: Include in the law alternative grounds for processing and solutions to protect privacy when obtaining meaningful is not practicable**

1. **If a new law were to add grounds for processing beyond consent, with privacy protective conditions, should it require organizations to seek to obtain consent in the first place, including through innovative models, before turning to other grounds?**

**Short Reply:**

Yes, despite existing practical challenges in obtaining meaningful consent from individuals concerned, it should remain a priority mechanism, but it should be supported by other obligations imposed on businesses pertaining to their governance, at time of design, and throughout the lifecycle of both the data they process as well as the technologies, AI systems or otherwise, they use to process such data.

**List of Recommendations**:

1. **Support consent model with mandatory demonstrable awareness and understanding tests or mechanisms**

2. **Augment receptivity of consent models by shifting from the traditional texts of terms and conditions towards multisensory tools (eg. video), based on the wide range of available digital technologies**

3. **Augment understanding of consent scope by imposing clear, simple and concise plain language to be used as an explicit compliance requirement.**

**Comments**:

One risk of eliminating the consent model mechanism is for less respectable businesses to increasingly and creatively chip and nibble at the flexible interpretation of processing obligations, which could gradually reduce the amount of legal and practical protection provided to individuals concerned. It may also provide incentive in implementing automated systems in "stealth mode", without the proper understanding and awareness of such individuals of the direct or indirect impact it may have on them. The lack of tangibility in most AI systems (subject to robotics), and the general sense of powerlessness individuals are feeling to date will be exacerbated into potential civil conflict and instability if mechanisms requiring their explicit and necessary action are de-prioritized within legislation.

The explicit consent mechanism could be supported by a mandatory awareness and understanding mechanism. Although concerns have been raised by marketing teams regarding the added friction of such consent or notification pop-ups or check-boxes, thereby reducing the seamless experience of the client or prospective client, such consent mechanisms could be reinforced with added requirements pertaining to "testing" the clarity of understanding of the user or individual concerned prior to giving his or her consent.

Exceptions to such mandatory consent, awareness and understanding mechanisms should remain limited to dire and exceptional circumstances, such as already existing and listed in PIPEDA.

Consideration should be given to impose and additional disclosure pertaining to transfer or processing of data by a third party through a business partnership relative to data about individual concerned. Such disclosure would clearly state that per the individual's personal data, business remains responsible for any non-compliance of business partners and other group undertakings towards such individual if the business provisions direct or indirect products and services with such individual or has collected, used, disclosed, processed, converted or otherwise shared such data with these business partners. Although the "unbreakable chain" liability is already stated in PIPEDA, the mention of such liability towards individual as a mandatory disclosure obligation at time of consent may reassure this individual as to being adequately compensated were there to be a breach of privacy or infringement of the privacy legislation, as well as force businesses to either restrict transfer of data to key business players with adequate safeguards in place, and put sufficient human resources, budget and protocols to adequately and continuously audit, monitor and verify that their business partners remain compliant with the law, to mitigate risk of their own liability.

Added practical guidance should be provided to such cooperation pertaining to investigation, verification, monitoring and enforcement against fraudulent or other criminal behaviour between private sector players, considering the societal and business opportunity isolation effect being put on a list of potential undesirables (ie. black list) may have on unsuspecting or good faith businesses and individuals. Liability in this respect is unclear, but businesses should be held to demonstrate that their officers are adequately trained, and that appropriate checklist and other technical and organizational measures -such as a human in the loop- are put in place to avoid detrimental mistakes to such individuals and business clients.

However, we are aware that obtaining a meaningful consent for all such personal or sensitive individual data, including the input data, "new" data created or inferred or converted by the AI systems, and the output data, will encounter technical and operational difficulties, as well as a certain complexity for the individual to provide consent for a potential evolving scope and processing of his data.

As such, the prospect of obtaining meaningful consent also comes into question when insights beyond the initially specified purpose are discovered. That is the basic premise behind utilizing machine learning is to discover latent patterns that are not evident to human analysis. Interpretability might be a tradeoff in using more complex models which might have a higher degree of accuracy. But that's not the case with the use of simpler models like binary decision trees that are highly understandable but lack predictive power in high dimensional, complex spaces. There is an emergent class of techniques, such as Local Interpretable Model Agnostic Explanations **(LIME)**[31], that allow for local analysis in the model that yield a meaningful compromise by approximating the behaviour of the complex model in a smaller context via a simpler model.

2. **Is it fair to consumers to create a system where, through the consent model, they would share the burden of authorizing AI versus one where the law would accept that consent is often not practical and other forms of protection must be found?**

**Short Reply**:

Yes, despite existing practical challenges in obtaining meaningful consent from individuals concerned, it should remain an existing mechanism within Canadian legislation.

**Comments:**

It is certain that there is a burden in requesting the end user or individual concerned to provide for a consent before using an AI system to process personal data. The cost

---

[31] RIBEIRO, Marco Tulio, OTHERS, "Why should I trust you?" Explaining the Predictions of any Classifier", https://arxiv.org/pdf/1602.04938v1.pdf

Montreal AI Ethics Institute                                    40

comes especially in limited time, where the individual is in a rush and wants to be provided with the product or service. It comes as well with a practice of ambiguous, complex, and long terms and conditions, written in "legalese" and not in clear and plain language, which requires a cognitive burden as well.

Finally, in practice, such terms and conditions or privacy policies are quite general in terms of defining the actual purposes and means as to how the personal data is being processed, and who has access to this information. Sections such as "your personal data may be discussed and share with our business partners concerned, relevant or necessary for the provisions of our products and services" or attempt to limit liability by referring end user to "such business partner's own privacy policy", are common, and have very little value in terms of assisting the individual concerned to develop sufficient awareness and understanding, and in providing a meaningful consent.

However, the consent mechanism is one of the few to provide some form of control over the collection, use, disclosure, processing, conversion and sharing of an individual's personal data. It could be augmented with, as proposed prior, awareness and understanding mechanisms (such as testing the user), prior to providing such consent. A legal obligation as to strictly forbid unclear, unnecessarily broad or potentially ambiguous legal language pertaining to all such disclosure obligations of business linked to privacy could also be considered to support the consent mechanism.

Finally, recent developments across the globe has shown that if the privacy and human rights framework pertaining to the implementation of new technologies such as AI is considered inadequate or distrusted by individuals, individuals have used their freedom of association and freedom of expression rights to publicly object against specific systems projects[32]. This may create civil dissension and instability. Considering these sentiments at time of policymaking is extremely important. An analogy can be made to the regulatory interventionism that was shown after the 2007-2008 financial crisis, where countries around the world put in place very stringent regulation against financial entities, for various reasons but also to demonstrate to the public that such a crisis would never happen again. We would suggest avoiding such a crisis pertaining to AI systems would be optimal, and as such, at least in a short-term, reviewable in a shorter span of time than 5 years, and perhaps a more stringent regulatory or legislative framework would be the answer to ensure the trust of individuals and the credibility of government and regulators as to ensure the importance of protecting human rights and freedoms.

---

[32]CRAWFORD, Kate, others, *AI NOW 2019 Report*, December 2019, 100 pages.

Montreal AI Ethics Institute                                                41

There have also been regulatory positioning that has been made at all levels of government, mostly on a local level, through a municipality decree, such as San Francisco's ban on facial-recognition systems[33], which was also recently considered by the city of Montreal[34]. This manner of dealing can take many analogies as to the recent federal legislation authorization the use, purchase and selling of cannabis, which was more stringently regulated in provinces and municipalities concerned. As such, other forms of protection which could support the consent mechanism are clear positions from the government against specific designs, developments, and deployment of systems.

We believe explicitly providing for this flexibility within the federal privacy legislation would be critical, as municipalities will be closest to the culture, concerns and expectations of their citizens to take positions per implementing specific technologies, that is, unless on the federal or provincial level, clear positions are taken against specific applications or purposes of automated systems. In any event, before implementing technologies on a city-wide or community-wide scale, serious consultation and feedback through either town hall events or other digital surveys or tools with the public and individuals which may be impacted by these systems should be done prior, to avoid civil unrest and group associations against the deployment of such technologies. What policymakers and cities must be prepared for is that if there is sufficient concern as to human rights and freedoms, it may first trigger civil uproar and in the event democratic dialogue appears out of reach from individuals following protests and other associations and lobbying, civil displacement into more human rights-centered or more protective communities or cities may ultimately become cause for concern[35].

It seems plausible that the public may expect in the short term that regulations provide for clear Go/No Go zones for specific AI technologies, unless sufficient guarantees are provided to them.

3. **Requiring consent implies organizations are able to define purposes for which they intend to use data with sufficient precision for the consent to be**

---

[33]CONGER, Katie, *San Francisco Bans Facial Recognition Technology*, 2019, https://www.nytimes.com/2019/05/14/us/facial-recognition-ban-san-francisco.html

[34]COLIN, Harris, *Montreal grapples with privacy concerns as more Canadian police forces use facial recognition*, CBC News, https://www.cbc.ca/news/canada/montreal/facial-recognition-artificial-intelligence-montreal-privacy-police-law-enforcement-1.5239892

[35]See note 22.

**meaningful. Are the various purposes inherent in AI processing sufficiently knowable so that they can be clearly explained to an individual at the time of collection in order for meaningful consent to be obtained?**

**Short Reply:**

Yes. Purpose limitation remains critical even for AI systems, to be identified at time of design, to be monitored and validated at time of development, training, testing, and to be meaningfully disclosed to individuals concerned at time of scaling or prior to use, to obtain consent.

**Comments:**

Design, development and deployment which may have the potential to directly or indirectly have an effect on an individual, in particular to the right of privacy but also other human rights and freedoms, should be strictly limited to an identified and intended purpose from the onset, as the foundational anchor and guidance for alignment for further programming. As such, an effort should be made at the time of design, for instance through a brainstorming session with a multidisciplinary team consisting of business product owners, IT officers and AI programmers, to identify all the potential and intended purposes for which it could be built, and then orient further developments based on such purposes relevant to individuals. Despite costs of obtaining consent of an individual for a new purpose identified in the later stages, this mechanism remains critical.

AI systems without specified purposes identified should remain in a strict sandboxed or lab environments within a corporation, with stricter requirements and safeguards, to ensure the protection of personal data of individuals, subject to their consent. This also applies to academic research and development labs as well, whether partnered with corporate sponsors, venture funds, or other special purpose vehicles. Individuals should maintain the freedom of refusing to let businesses further develop their R&D teams and analytics and systems capacities with their personal data. Guarantees as to sufficient conversion, anonymization or differential privacy measures to de-identify data inputted through this R&D could be disclosed at time of obtaining consent. But regardless, individuals should be able to say no, including when analytics are founded on an individual's browsing history or searches on search engines. Individuals should maintain the right to refuse analytics to be made using their personal data.

Per the latter, caution should however be taken since some business models have made possible through the democratization of access to different digital tools and access to information on the basis on being able to generate sufficient profits in the personal data analytics linked to the use of their products and services. The democratization and access is noteworthy and could be considered a social good. The importance here remains to strike an adequate balance, and it should remain possible to determine to which extent such analytics should be permitted. This may require a country-wide positioning, with specific examples founded on regulatory decisions, regulatory gaps, and citizens' sentiment and expectations. An interesting obligation that could be further explored is that of forcing such business models to provide for a mandatory alternative, analytics-free, option to the individual, as an alternative to the analytics use case. Evidently, such pecuniary options should not be excessively demanding as to effectively bar a majority to opt for such an alternative.

Finally, relative to meaningful consent, we are aware that there may not be any real efficient way that has been demonstrated so far that ensure meaningfulness even for simple products let alone complex ones that use AI. Additionally, the public at large is desensitized to consent forms and more so do not have enough time to process it in a meaningful manner for all the services that they interact with[36]. Secondly, there might not be alternatives available which makes consent a moot point, often the case with digital monopolies. However, there are alternative formats and innovation ways to obtain meaningful consent, especially if lawyers and businesses shift away from textual terms and conditions and towards multi sensory tools, such as videos, to increase reception and understanding of how personal data is being processed by business concerned. In any event, additional mandatory mechanisms as proposed in this response should be considered to augment protection of individuals.

4. **Should consent be reserved for situations where purposes are clear and directly relevant to a service, leaving certain situations to be governed by other grounds? In your view, what are the situations that should be governed by other grounds?**

**Short Reply:**

We consider the question could have been drafted differently, in a manner that would clearly state that consent is not reserved to specific decisions or situations, but is a mandatory obligation applied generally to all situations, unless specific situations should permit processing of personal data without consent but in dire circumstances. Such scenarios such as already listed within PIPEDA appear reasonable to us.

---

[36]HERN, Alex, "*I read all the small print on the internet*", The Guardian, 2015, https://www.theguardian.com/technology/2015/jun/15/i-read-all-the-small-print-on-the-internet

**Comments:**

We have already provided comments above pertaining to additional practical guidance and obligations in the method and manner of exchanging or processing data between private sector businesses in the context of investigating, verifying or enforcing list of "undesirables" associated with real or reasonably suspicious activities of fraud or other criminal offence, to avoid detrimental or erroneous listing.

From our perspective, the Canadian legal ecosystem pertaining to data is structured in siloed, specific sectors, which could be prone to deficiencies and gaps detrimental to Canadian individuals. For instance, PIPEDA only covers activities in the private sector, and regulates federal businesses and entities at all times, but may defer to provincial legislation for provincial businesses and entities if such legislation is found to be essentially similar to the federal one. Some provinces either apply PIPEDA, have provincial general legislation encompassing all types of data for the private sector, and then a public sector privacy legislation, and then other provinces have provided for specific legislation pertaining to healthcare, genetic or biometric data. Finally, national security or military applications fall out of scope of all of these. Additional transparency pertaining to national security and military applications, subject to tactical or strategic decisions made in this respect, would be greatly recommended. We consider that further legal protection is required pertaining to healthcare, genetic and biometric data, to ensure adequate protection.

**Employee data:** We also consider that further consideration should be given pertaining to employee data. PIPEDA is more lax in this regard, however, we have growing concerns that employees may feel a "chilling effect" by having their activities and tasks constantly monitored and micro-managed by AI systems to score their performance with potentially excessive targets, thereby exacerbating risks of anxiety and burn-out[37]. Today, human resources divisions are implementing systems to take critical hiring and firing decisions based on such inferences made by AI systems which are prone to evident limitations in this respect. We consider that the laxness PIPEDA shows in this respect should be resolved, to provide additional guidance as to how and for which purposes AI systems may use and make decisions based on employee data. Otherwise, as done in other jurisdictions, we predict labour syndicates may find in AI a new threat to excessively aggregating power in the hands of the employer[38].

---

[37] DZIEZA, Josh, "How Hard will the robots make us work?", The Verge, 2020, https://www.theverge.com/2020/2/27/21155254/automation-robots-unemployment-jobs-vs-human-google-amazon

In addition, PIPEDA does not apply to public sector entities, which are held to other legislative standards. We recommend public sector legislation to provide for more stringent safeguards than that of the private sector, due to the sensitivity and volumes of personal data they hold. We also recommend that PIPEDA explicitly states that private sector businesses which enter into partnerships with government must both comply to PIPEDA and public sector standards (a similar obligation to be drafted within public data legislation), and in the event of potential or unforeseen conflict, will be held to the highest standard. This is to ensure there are no gaps in compliance and in enforcement for public-private partnerships, and to avoid a "too connected to fail" problem[39].

5. **How should any new grounds for processing in PIPEDA be framed: as socially beneficial purposes (where the public interest clearly outweighs privacy incursions) or more broadly, such as the GDPR's legitimate interests (which includes legitimate commercial interests)?**

**Short Reply:**

Based on our experience, the concept of "legitimate interests" as a broadly defined concept in GDPR, despite examples and guidelines provided, is not particularly well perceived or received by the public, due to the greyness or potential ambiguity linked to such concept.

**Comments:**

We consider PIPEDA took an essentially similar but clearer approach in identifying the key exceptional circumstances for which businesses may collect, use or otherwise disclose data pertaining to individuals to enforce their civil rights associated with the ordinary course of business as reasonably available to them. We consider the listing approach is optimal, in terms of providing clear margins and clearly acceptable behaviours. As such, including the legitimate interests concept within Canadian legal framework does not hit us as having plus-value, and we believe sufficient arguments

---

[38]See AI Now 2019, note 22.
[39]ARNER, Douglas, OTHERS, *The Dark Side of Digital Financial Transformation: The New Risks of Fintech and the Rise of TechRisk*, University of Hong Kong Faculty of Law Research Paper No. 2019/112, 37 pages,p. 5.

can be provided to ensure Canadian law meets the GDPR's adequacy test in this respect.

Evidently, there may be an argument to state that being able to use data collected from individuals to develop AI systems internally or through partnerships would be of tremendous interest to business, and which could be justified as a potential social good interest for individuals as well. This may also provide an interesting incentive for supporting innovation. If such "legitimate commercial interest" is to be agreed upon, we recommend it to be explicitly added to the law, but with clear, formal and stringent requirements pertaining to de-identification technical measures, and organizational safeguards.

6. **What are your views on adopting incentives that would encourage meaningful consent models for use of personal information for business innovation?**

**Short Reply:**

These incentives are absolutely necessary, considering the potential costs of organizational and technical restructuration associated with augmenting meaningful consent models.

**Comments:**

Fiscal incentives pertaining to AI research and developments and other grants are already made available, but fiscal incentives pertaining to augmenting cybersecurity structures and privacy and human rights models would be extremely interesting and would nudge businesses in the right direction.

In a context where it cannot be guaranteed privacy breaches will not happen, and as we fall into risk management protocols, other incentives could include, at time of finding liability for non-compliance, as well as at time of determining amount of administrative penalties and applicable measures, considering explicitly lessening these based on the sophisticated nature, degree and level of technical and organizational measures put in place to ensure meaningful consent and against privacy breaches. This would be identified as a factor to be weighed for liability and amount of damages. Another weighable factor in these circumstances could be the level of privacy literacy and other tools provided to the public to augment their awareness, understanding, competence and education about their privacy and consent rights, in particular pertaining to automated decision-making systems. If it can be shown that the business takes its social responsibility seriously through this coaching and training, it can potentially lessen liability level and damages.

**Proposal 8: Establish rules that allow for flexibility in using information that has been rendered non-identifiable, while ensuring there are enhanced measures to protect against re-identification**

1. **What could be the role of de-identification or other comparable state of the art techniques (synthetic data, differential privacy, etc.) in achieving both legitimate commercial interests and protection of privacy?**

**Short Reply:**

De-identification should be identified as the corporate governance norm for protection of privacy, to be graduated against type and sensitivity of data, unless specific consent by individual for business to use personal data is provided. Furthermore, where data has been identified and business has interest in using or processing such data to improve its own AI systems or analytics, assurances about the appropriateness and degree of security of de-identification techniques should be disclosed to the individual concerned.

**Comments:**

Based on our knowledge, differential privacy at the moment appears to offer the best alternative that is computationally tractable and provides mathematical guarantees that are tunable according to the use case, although there are discussions that there may be some limitations to such a technique pertaining to deep-learning models and neural networks which should be further explored.

Synthetic data creates an interesting dilemma, as they could be a way to provide alternate sources of data to train machine learning systems but their adequacy in providing sufficient richness is still in question. Generative Adversarial Networks (GANs) provide a potential approach to generate synthetic data distributions that might be useful but there is more research required to verify the efficacy of such an approach.

Homomorphic encryption offers another alternative where data is obfuscated via encryption that preserves the mathematical properties of the dataset on aggregate which is useful for performing computations and learning statistical information from it. Problem with it remains that the computations are intractable for major computations though the experiment done by Numerai provides an example where this was applied to financial data analysis with success.

Montreal AI Ethics Institute                    49

Thus, we can protect commercial interests with a modicum of privacy protection based on the structure of the system and implementation of one or more above techniques in varying combinations as suited to the sensitivity of the application.

When talking about a reasonable risk of re-identification, consideration need be given as to legally define what will be considered "reasonable" as a test to ensure technical and organization measures are considered appropriately and sufficiently compliant, as well as efficient protection of individuals. The test should also be drafted in a manner that can adequately account for future evolution of technology which may reveal new de-identification and re-identification techniques.

The Japanese law provides an interesting example though it is uncertain whether it holds well in the world of machine learning where patterns can be drawn and applied retroactively against other datasets that are publicly available, and which could lead to a "mosaic effect", effectively re-identifying individual through data associations, correlations or inferences.

The risk of financial fines or imprisonment is insufficient for particularly sensitive data, for example biometric data, which, once re-identified and potentially stolen, causes irreplaceable damage, because these things cannot be changed like an address or even SIN. Also for DNA or genome data, it also affects related individuals beyond those that were a part of the compromised dataset. As such, more stringent mechanisms must be imposed for more sensitive data, graduated perhaps on the basis on whether or not such data has a permanent or temporarily sensitive nature.

2. **Which PIPEDA principles would be subject to exceptions or relaxation?**

No comments.

3. **What could be enhanced measures under a reformed Act to prevent re-identification?**

**Short Reply:**

Please see question 1 above.

Montreal AI Ethics Institute　　　　50

> **Comments:**
>
> Enhanced measures should be linked with cybersecurity measures in this respect, to ensure defence in depth, both as organizational and technical measures. For instance, NIST cybersecurity guidelines[40] can provide for key governance safeguards against re-identification.
>
> In addition, differential privacy provides for strong protections against re-identification, subject to some limitations. Finally, caution is advised pertaining to rendering data publicly available since this may provide sufficient association, correlation or inference between different de-personalized data using the mosaic effect.

---

[40] NIST Cybersecurity, https://www.nist.gov/topics/cybersecurity

**Proposal 9: Require organizations to ensure data and algorithmic traceability, including in relation to datasets, processes and decisions made during the AI system lifecycle**

**Is data traceability necessary, in an AI context, to ensure compliance with principles of data accuracy, transparency, access and correction and accountability, or are there other effective ways to achieve meaningful compliance with these principles?**

**Short Reply:**

Yes, to ensure adequate accountability, data traceability is necessary to ensure compliance, subject to such limits pertaining to avoiding risk of re-identification by using such records pertaining to data traceability, as well as avoiding identifying weaknesses and sources which could be gamed upon by malicious third parties.

**Comments:**

We are of the view that meaningful compliance may be hard even given data traceability requirements. Additionally, it may be really hard to enforce and define this requirement uniformly across firms.

Data lineage and data provenance are indeed noble aims but they are much harder to require, particularly concerns arising from companies who would hesitate in making public their data pipelines and the steps that they take to transform and combine data because that would erode a significant part of their advantages. A trivial example would be that models can be sourced from a largely accepted state of the art literature and open-source repositories, for example on GitHub, in the field and would vary less than the actual combination and utilization of the data pipeline along with the system configuration including the models. Thus, a higher portion of the value add of the firm will reside in their data stages rather than the model stages though this is hard to delineate quantitatively and thus hard to enforce and offers a legal gray zone that companies can use to get out of having to reveal internal working details.

The major risk seen with the *Algorithmic Accountability Act* is that it can lead to a gaming of the audit ecosystem as happens in the world of finance again and again. Plus, each of these will require high degrees of specialization because of the vast variation in different ML domains like computer vision, NLP, etc which might be hard to unify under a single standard.

The PwC suggestion seems insufficient because without a standardized labeling process amongst other steps, the evaluation across a range of firms by the same auditing company will lead to widely differing results.

The reasonable notice requirement is essentially ineffective from a technical perspective because there is no public benchmark against which the consumers could compare whether they were mischaracterized, even so, it may be impossible to tell which factors in the input datasets and it's interaction with the model led to the decision and how the individual could get a fairer result from the system when there is a lack of explainability and interpretability of how a complex model arrives at a particular result.

**Proposal 10: Mandate demonstrable accountability for the development and implementation of AI processing**

1. **Would enhanced measures such as those as we propose (record-keeping, third party audits, proactive inspections by the OPC) be effective means to ensure demonstrable accountability on the part of organizations?**

**Short Reply**:

Yes they would but as pointed out above, the demonstration does not entirely achieve what the law is seeking to achieve.

**Comments:**

Pertaining to the proposal of public filing, for complex systems that might be inscrutable, there is very little that we can do to verify if the public filing is accurate and covers all aspects of the system and captures all downstream effects which are even harder to capture.

In respect of the proposal that auditors could be fined for signing off on systems that don't meet requirements (no legal or contractual immunity), our comment is that similar accountability is expected in the financial world, but fraud continues to happen, where fines' threshold are still out of sync with the gains to be made by flouting laws. The GDPR approach in identifying a percentage of turnover growth is found to be more effective than a specific amount.

In addition, it seems to be hard to detect in the first place that the company meets the requirements in letter but not in spirit. This could lead to signing off on the systems and that seeming to be right as well and detection of flouting only happening in the case of a whistleblower under the assumption that the systems are sufficiently complex that they evade meaningful analysis.

The ITLA recommendation is interesting for keeping the blame allocation on the human behind the system. This is in line with the GDPR. The only specification this needs is on how to reconcile emergent behaviour of a system when it learns and adapts by interacting with new, real world data. One proposal to counter this is to have the explicit requirement to have guardrails that constrain the behaviour of the system within predetermined thresholds. Additionally, it would also require instrumentation to monitor shifts in data distributions compared to what the system was trained on.

Specific to the proactive inspections power considered for the regulator, we consider that this may provide additional incentive for businesses to ensure ongoing data protection compliance, instead of a regulator having to wait for an individual to file a complaint or to develop reasonable suspicion or doubt prior to inspection or investigation powers being available.

However, due consideration will need to be given to ensure an equitable and fair due process in the context where through proactive inspection, regulator discovers and identifies alleged non-compliance which may have adverse consequences on the business concerned, eventually leading toward administrative fines or civil recourse or other regulatory measures imposed. Businesses will need available procedural protections, as well as the right to be heard and to challenge the findings of the regulator with all such administrative law defences available to it.

A graduated approach could be considered in this respect, where after a proactive inspective, regulator provides with preliminary findings and recommendations on which technical and organizational measures it would expect to see implemented by business in a reasonable period of time, without necessarily leading to damages. However, where the period of time provided expires, the regulator would have obtained reasonable suspicion of infringement being committed or having been committed, and then would go through the regular investigation process, with all due procedural protections. The proactive inspection could be associated with a compliance sandbox service provided by the regulator concerned, to assist the business in transitioning towards the required threshold or standard of data protection.

Additionally, in the event that an administrative tribunal specific to data protection were to be appointed within a specific legislation, all such traditional procedural safeguards would have to be made available to defendants, such as the right to be heard and to challenge the findings, as well as the right to appeal to a superior court where an error at law is alleged.

Furthermore, consideration need be given as to provide explicitly as to whether there are civil recourses and remedies available under such specific legislation and if so, whether or not it provides for a "complete code" procedure and defence as against all other civil claims or recourse which would otherwise be available to individual under a civil procedural code or at common law. In the short-term, we recommend that such complete code defence be explicitly put to rest, to ensure that individual affected may at will go through the traditional civil system and through the data protection legislation, to avoid AI systems accountability falling in gaps between different legal regimes.

In addition, in terms of non-compliance, some other regulators were provided with the discretion of avoiding publishing the name of non-compliant business in their administrative decision, based on objective criteria, where it is found that publication would unduly damage the business to an unreasonable degree. This discretion is normally not well perceived or accepted by the public, since it goes against the transparency principle of publishing judicial decisions in Canada, however, it is an available tool which may foster further cooperation between regulator and businesses concerned as well. We do not recommend publication anonymity of administrative decisions made by regulatory administrative tribunal, subject to some exceptions pertaining to clear dire circumstances, one can look at the GDPR for examples. However, anonymity at the time of proactive inspections may be a good mechanism to alleviate procedural burden and substantive administrative law safeguards.

Finally, consideration need be given to include criminal sanctions in a data protection legislation or not. We understand that some sections particular to data protection and unauthorized access are already included within the Criminal Code. In practice, even if there is explicit opportunity to order criminal sanctions against designated individuals in some regulatory framework, it is very rare to see a regulator press for such. Understandably, higher procedural safeguards and constitutional rights would have to be made available to the defendant. Caution is advised to avoid duplicating sanctions (ie. *ne bis in idem*).

2. **What are the implementation considerations for the various measures identified?**

See above.

3. **What additional measures should be put in place to ensure that humans remain accountable for AI decisions?**

**Reply:**

As already set out by existing legislation, the designated data protection officer **(DPO)** or chief privacy officer **(CPO)** could be mandated with ensuring that technical and organizational measures have been taken for AI systems to prevent or mitigate human rights and privacy infringements and to be processed in compliance with law. If DPO is unable to demonstrate sufficient technical knowledge or expertise pertaining to such systems, then a "responsible and safe systems" officer **(RSSO)** could be designated in addition to the DPO.

Finally, similar escalation procedures, such as notification, reporting and regular reviews mechanism from and to an executive board, to ensure top-to-bottom awareness and accountability as well as role modelling in corporate values to such effect, would also be recommended for optimal governance.

**Proposal 11: Empower the OPC to issue binding orders and financial penalties to organizations for non-compliance with the law**

1. **Do you agree that in order for AI to be implemented in respect of privacy and human rights, organizations need to be subject to enforceable penalties for non-compliance with the law?**

**Short Reply:**

Yes.

**Comments:**

In fact for marginalized populations to feel comfortable in utilizing these systems and being subjected to decisions from such systems they need to be confident that there is a recourse system available to them that protects their fundamental rights. Additionally, for inclusive participation and encouraging higher rates of participation from those that are on the margins we need to have trust be evoked in the interactions with the system. It is recommended to provide swift enforcement and penalties as a way of building trust. Finally, there needs to be a build up of civic, nonprofit agencies that can assist and educate consumers on how best to utilize such provisions which would otherwise let slip too many vulnerable populations because the burden would lie on the consumer to understand when a problem has occurred.

In terms of an independent regulator with enforceable powers and how to structure such, we suggest reading our response to the Australian Human Rights Commission about the Responsible AI regulator they were proposing for appointment[41].

2. **Are there additional or alternative measures that could achieve the same objectives?**

**Short Reply:**

See comments under Proposal 10.

---

[41]GUPTA, Abhishek, SNYDER CARON, Mirka, "Response to the AHRC and WEF regarding Responsible Innovation in AI", Montreal AI Ethics Institute, 25 pages.

**Comments:**

Consideration should be given to provide for a presumption, or shifting the burden of proof, both under regulatory recourse and civil recourse of applicant, unto the defendant business concerned to help alleviate costs and lack of expertise of individual concerned requesting remedy for breach or infringement.

Consideration should also be given to formally appointing an independent AI systems auditor for compliance purposes.

Enforcing strict adherence to the law will be crucial to setting the stage for how firms behave and in ensuring that they follow the law under a regime where it is infeasible to check on the entire scope of operations for every firm (even under external auditing where we suspect something similar to the statement of applicability problem might arise. The SoA or statement of applicability problem alludes to the issue that arises commonly when a certification is issued for a system, and the SoA is not made public, people interacting with the system assume that the entire system is certified and adheres to the standard which is misleading because the SoA actually illustrates the precise scope of the system that was certified and found to meet the standard which is almost always not the entire system but just a subset of the system. )

**SCHEDULE 1**
**PUBLIC FEEDBACK FROM MAIEI WORKSHOPS ON PRIVACY LEGISLATIONS AMENDMENTS RELATIVE TO ARTIFICIAL INTELLIGENCE**

Dates: 24[42] and 27[43] February, 2020
Editor: Mirka Snyder Caron
Authors: Montrealers
Total # registrants: 73.

**Topic 1.: AI in privacy law: Should AI (artificial intelligence) be governed by the same rules as other forms of processing, (ie. no need of defining AI and maintaining a technologically neutral law), or should certain rules be limited to AI (artificial intelligence) due to its specific risks to privacy and, consequently, to other human rights?**

**Short Reply**:

Defining AI at law is difficult and may fall short of preventing actors from circumventing the definition to avoid regulation. However, a solely technologically neutral law is not the answer either.

As such, specific sections focusing on technical and organizational processing of data, as well as increased transparency, and greater control of the data to be shifted to the user, would be technology-specific sections to legislate about for appropriate protection and guidance to actors and stakeholders concerned.

**Comments**:

There are unique characteristics of AI, which go beyond automation, and by which processing of data is "scaled up" and may have effects when applied to decision-making processes. In terms of determining whether it is possible to define AI, the group identified that attempts to circumvent AI-specific regulations were made by introducing a "human in the loop", and this should be prevented.

---

[42]https://www.eventbrite.ca/e/ai-ethics-quebec-and-canada-ai-privacy-legislations-part-1-tickets-95677119841
[43]https://www.eventbrite.ca/e/ai-ethics-quebec-and-canada-ai-privacy-legislations-part-2-tickets-95689171889

It was concluded that an AI definition would be very vague, as it is not easy to define automation in a way which would restrict privacy laws to processing AI algorithms.

Instead, the group explored what would be necessary to define or regulate. They identified bias and discrimination as a priority. They identified it was mostly the dataset that is biased and not the algorithm, although the algorithm can significantly reinforce "bias" as processing is "scaled up" without human intervention, which problems go against merely having a technologically neutral law.

Questions arose in terms of the method to prevent bias in datasets by introducing specific privacy laws for AI, and whether the target should move away from AI specifically and more towards the methods or processes of collection and structure the data goes through.

Additionally, there are a lot of actors and stakeholders in the private sector using AI technologies, and the group concluded that actors such as data brokers and processors should be subject to higher security standards, which goes against the precepts of technologically neutral law.

Users should be made aware as to why and how their data is used by AI technologies. Different important factors or conditions were identified, mainly:

- The readability is important
- Education to raise awareness on these disclosures
- Focus should be on protection of all personal data and not just as used by AI, which required more transparency and control, in a manner that goes against the precepts of technologically neutral law.

Concerns were raised as to whether the logic of an AI system could be adequately explained to a layperson, and in conclusion, it was found that due to increased risk in standardizing discrimination, users should have more control through transparency, and higher security standards to be implemented throughout the data lifecycle.

Some specific use cases of AI were identified as most important:

- Facial recognition
- social credit scores

- law enforcement

The conclusion was that it is difficult to define AI at law with a sufficiently broad scope, but that a technologically neutral law would be risky, considering the risk present technologies have in reinforcing bias and illegal discrimination, and affecting human rights and freedoms. As such, specific sections imposing greater transparency and control to the user over the data were identified as primary areas of concern.

**Topic 2.: Right to object: Should PIPEDA (Personal Information Protection and Electronic Documents Act) include a right to object as framed in this proposal? If so, what should be the relevant parameters and conditions for its application?**

**Short reply:**

**Yes,** there should be a right to object, and/or another concept such as a right to negotiate an automated decision with a human, and/or a right to a reasonable inference made the automated agent.

**Parameters and conditions for this right are as follows:**

- Such right must be made explicit by law
- Transparent disclosure to individual that the decision was made by a human or by an automated agent
- To be weighed against the hierarchy in types of decisions made, in other words:
    - Whether the decision has, or has the potential to, have a direct impact or effect on individuals
    - Whether the decision has, or has the potential to, have an indirect impact of effect on individuals
    - Whether the decision has no, or has no potential to, have any impact or effect on individuals
- The intensity of the impact on the individual, community, or collectivity
- The scale of the impact on a same individual, community, or collectivity
- The type and sensitivity of data
- Disclosure of objective and subjective criteria programmed in the automated agent, and the right to contest the input of subjective data which may negatively affect the result of the decision for an individual
- To be weighed against the types of impact, the following being non-exhaustive examples:

- Whether the decision has, or has the potential to, have an impact on the rights and freedoms of individuals and/or communities
- Whether the decision has, or has the potential to, have an impact on the health and well-being of individuals and/or communities
- Whether the decision has, or has the potential to, have an impact on the economic interests of individuals, communities and/or entities
- Whether the decision has, or has the potential to, have an impact on the sustainability of an ecosystem, social, environmental or otherwise.
- The efficiency of anti-discrimination laws

**Comments:**

The general answer to the above mentioned question was **yes,** the individual should be able to object to a specific business decision made about him by an automated agent and which may have an effect on him. Other concepts of such rights were discussed, such as the "right to negotiate" an automated decision with a human, as well as the right to a reasonable inference by such automated agent, as in how much would be considered "reasonable" or not, and also how much arbitrariness or "freedom" in the business criteria ought be provided to businesses to build their models.

Greater transparency was requested for identifying whether a decision was made by a human or an automated agent.

It was also proposed that due to the lack of individual's power against businesses, a legislative framework was needed to level the playing field and balance powers between businesses and individuals.

Considerations were given into determining the hierarchy in the types of decisions that could be made. For instance, some decisions have a direct impact or effect on individuals, while others do not or may impact the individuals indirectly. Such impact may also vary in intensity for the same individual, and in scale for the same community.

Concerns were raised as to determine whether or not the criteria embedded within the automated agent was objective or subjective, and how to mitigate or prevent risks associated with detrimental or discriminatory subjectivity.

The hierarchy of the criticality to such right to object would have to be generally considered in conjunction with the following parameters to be weighed:

- the rights of individuals and/or communities (individual good AND collective good)
- the health and well-being of individuals and/or communities
- the economic interests of individuals, entities, and/or communities
- the ongoing sustainability of the ecosystem.

Furthermore, the right to object would be hierarchized based on the type of data that is being analyzed, or in other words, the sensitivity of the data. However, it was considered that alone such criteria was not good, since even non traditional sensitive data, one can reconstruct very precise profiles.

Another consideration was that of the importance of effective anti-discrimination laws, as part of the protection of human rights and freedoms of individuals, including that of the right to privacy. Arguments were raised in saying that to avoid gaps or competencies overlaps between different legislative frameworks and regulatory mandates, the potential opportunity of overhauling the siloed privacy approach and create an effective ecosystem of data protection with a broader legislative framework than that focused solely on the right to privacy, would be more efficient, but would require major restructuring of existing regulatory entities.

The right to object was linked with the right to explainability, by saying that to accept a decision, there needs to be sufficient trust in the system. It did not seem at the time of the discussion that there was much trust in the manner, method and governance businesses were going to design and deploy automated agents without more stringent legal requirements.

As such, the right to object was linked to protecting and preserving the ability to think, to make personal decisions (ie. right to self-determination), and to be creative which, due to the potential loss of flexibility in the automation of decision-making systems, this could be negatively impacted. For instance, AI based decision-making tools can introduce at least 2 kinds of bias, data bias and cognitive bias, cognitive bias being defined as absolute reliance on decisions made by the automated agent, and in such case you eventually lose critical cognitive reflexes (this is also linked to an inertia and over reliance biases).

**Topic 3.: Right to explainability: If incorporated, what should it entail? And would enhanced transparency measures significantly improve privacy protection, or would more traditional measures suffice, such as audits and other enforcement actions by regulators?**

**Short reply:**
**Yes,** it should be incorporated.

3 different categories of factors to be met:
-Factors pertaining to data
-Factors pertaining to the result generated by the automated agent
-Factors pertaining to cybersecurity risks

**1) Factors to be disclosed and questions to be answered to ensure explainability were identified as follows:**

- What data is being used?
- What data specifically is necessary for the identified purpose?
- Why is the data needed?
- Re-disclosure when purpose is changed/expanded -> new consent required
- How is the data used?
- How is the data accessed?
- From where is the data being accessed?
- Who has access to the data?

**2) Factors explaining how the output of algorithmic data processing were generated:**

- Logic of decision-making
- Decision criteria
- How is data displayed (aggregated or individualized)?
- Explanation of access rights and consent protections
- How can users have access to the data collected on them?
- How can users retract or limit their consent?
- Other consequences on rights and interests
- Explanation of safeguards against bias in algorithmic decision-making -> would necessarily be a continuous improvement, even disclosure of safeguards would always need to be updated
- How was the data set collected?

- What variables are used in the algorithm?
- What variables may serve as proxies for identity-markers that may be a ground for discrimination?
- What efforts are being made to track improvements in rooting out bias?

3) **Factors to respond to cybersecurity concerns**:

- What is the company doing to keep personal data safe?
- What is the level of "trackability", based on provenance, lineage and vulnerability?

**Yes,** enhanced transparency measures will improve protection, **and no,** traditional measures will not suffice and present audit and enforcement measures need to be **more deterrent and credible.**

**Comments:**

Enhanced transparency will ensure more trust between individuals and businesses and, an incentive for more responsible and efficient privacy governance by businesses concerned.

Considering cybersecurity issues and implementation of automated agents, present developments require enhanced transparency and enforcement mechanisms by governmental actors.

It appears that pertaining to cybersecurity and human rights risk of automated agents, businesses would be held to obligation of means rather than obligation of results. It is expected that the government would commit itself in making the best efforts to safeguard against bias, and protect against data breaches, etc., and a similar commitment is expected from businesses as well. Realistically, an absolute commitment to precluding bias or data breaches from occurring in the first place is not possible, since the technology evolves too fast to be able to commit to an obligation of results.

Measures of transparency were found necessary, as findings were that the current traditional measures do not suffice. It was proposed to put in place an enhanced OPC (Office of Privacy Commissioner).

A possible recommendation was to put in place a specific governmental agency responsible for evaluating new AI technologies and applications in areas of public interest, in other words, having an AI-specific version of Pest Management Regulatory Agency **(PMRA)** and other regulatory agencies.

Responsibilities of such AI regulatory agency would be to oversee the enforcement and compliance with transparency measures. There would be a need to have 2 different implementation models, one for the private sector, and another for the public sector. Other quality control and expert inspection would be available to validate accuracy of models and to efficiently track efforts in improving such accuracy.

To address and oversee safeguards implemented against bias, it was proposed that there should be tracking of such obligation of means, and a disclosure of the algorithmic design when the private company has a public sector mandate.

Concerns were raised pertaining to the use of personal data without consent. It was proposed that there should be a generalization of AI regulations beyond just AI applications: as such, the regulation would cover any use of personal data. It was suggested that public institutions which are not held to a duty to disclose the use made of personal data were typically the ones hoarding the most personal data.

It was also proposed that for adequacy reasons and other reasons, there should be comparative brainstorming sessions made with the *EU Guidelines on Artificial Intelligence and Data Protection,* in particular pertaining to the right to obtain information on the reasoning underlying AI data processing operations applied to them, and the consequences of such reasoning.

**Topic 4.- Privacy by Design & Mandatory testing: Should Privacy by Design be a legal requirement under PIPEDA? Would it be feasible or desirable to create an obligation for manufacturers to test AI (artificial intelligence) products and procedures for privacy and human rights impacts as a precondition of access to the market?**

**Short reply:**
**Yes,** privacy by design should be made a mandatory legal requirement under PIPEDA.
**Yes**, it is both feasible and desirable to impose such an obligation.

**Proposition:**

- Prior to testing, establishing:
    - Ethical expert within the company (having both tech & ethics expertise), modelled on resident medical ethicist
    - Establish a self-regulatory professional board of experts (diversity in knowledge), modelled on existing professional bodies (medical, lawyers, banks, etc.): they would oversee the testing of AI systems.
    - Apply for a license for each project
        - License differs according to the sensitivity of the data & vulnerability of the audience

Implementing a point system like the driver's license (incentives, etc.)

**During testing:**

- proceed in an iterative instead of adversarial way(questions; incentives; etc.)
- Building in a feedback system (for explainable AI)
- Pushing a disclaimer at the release

**General framework to ensure efficiency:**

- Give more power to the Privacy Commissioner:
    - (General recommendation) Selected by multi-party committee in Parliament instead of the party in power
- Ongoing consent of the user/audience
- Compliance with ethics as a requirement for obtaining a tax break or fiscal incentive

- Community service if not compliant
- Each project: ongoing ethics training for tech experts. Like getting an ethics approval. (Makes sure that we are in sync with technology)

**Comments**:

**Privacy by Design:**
Privacy by Design was emphasized as being the obligation to include privacy protections into the technologies themselves. It was found that the difficulty lied not in implementing, but ensuring efficient auditing and monitoring of such. In terms of building in privacy, the example of the feedback system like XAI in the States was identified, to explain decisions.

It was brought up that there are often public statements made by businesses assuring that "Your privacy is important to us". However, despite terms and conditions and privacy policies, at times it is still not possible to know how the data is used. Additional details ought to be provided to tell the consumer or individual concerned the way to find out how it is used, and if there is a breach, a way to know how the breach happened.

**Mandatory obligation for testing AI prior to access to market:**

It was concluded that manufacturers should have a mandatory obligation to test their AI systems against privacy and human rights impacts before gaining access to the market or marketing their products or services. There was awareness at how difficult this could be, and what would be the best way to ensure this be done.

Another difficulty was identified, namely, that in practice, tech designed wait until the problem happens because the technology is not designed to foresee problems, but to deal with them as they come. The proposal to remedy this issue was to put in place a self-regulating board, with sanctions and licences, such as a professional order that can be found for doctors, engineers and lawyers. In such a way, a certain degree of professional investment is required to maintain a certain level of ethics and professional conduct.

**Augmenting deontology and ethics obligations of engineers and programmers:**

It was perceived that engineers' deontology code, as compared to doctors or lawyers, seemed to be the weakest. Furthermore, many programmers to date have learnt to

Montreal AI Ethics Institute                                      69

code and program without any standard academic curriculum or licensed training, which would ensure that adequate training and testing be made pertaining to ethical and responsible AI design and development, including that about preventing and mitigating impacts on privacy and other human rights.

Various examples were provided from the practice and field of medicine. It was found that there may be a lack of teaching the people who deal with the data about 1) the tech literacy issues and 2) the implication of using such technology surrounding automated and machine-learning models and ensuring privacy protection when using such sensitive health, biometric and genetic data.

It was proposed that within hospitals and other research labs and manufacturing labs, there ought to be a medical ethics and/or bioethics expert, which was defined as someone who would be an expert with training in applied ethics & in computer science, in other words, someone who know how to ask the right questions, as well as someone who understands how people code, so that such expert may process questions to coders. It was deemed mandatory that such an expert was to be a coding expert.

**General framework for testing AI:**

The proposal stated above was then generalized to recommend there to be an ethical expert within a company to assess the impact and oversee the testing of the AI, one that would have the expertise in AI & ethics, tech & philosophy. To create an analogy, the same way a data protection officer needs to be designated for a company, a similar title and position would be designated for AI systems.

Another proposal was to implement a board of experts to oversee such testing, which would be provided with the explanation of how the AI was built and how it worked, how it was structured, and to disclose this to the public, without necessarily revealing the intellectual property protected aspects of it.

Additional organizational measures could be taken by training the board members to augment their technical literacy, and to regularly test such knowledge and maintain it up to date.

Particular to the testing, it was proposed that it ought to follow a step-by-step process as the board members or other experts go through the approval of projects, favouring the iterative process over and adversarial design in technologies. Furthermore, the process should take into consideration whether the project would take in a particular sector of the market or economy, identified as the 3Ps: Private sector, Public sector, and Plural (combined partnership).

It was also suggested that the regulatory body in charge should be diverse in terms of knowledge, and have an explicit obligation to ensure such diversity. More power should be delegated to the Privacy Commissioner in terms of a reward and penalty system, akin to that of a driver's license points and score, and to treat responsible AI ethics, and privacy ethics as a baseline to such scoring.

**General conditions for such system were identified as follows:**

- penalties to be proportional to the repercussions implied (principle of proportionality)
- punishment should not be only financial but should expand to such bans against ongoing or future commercial projects and/or governmental relations.
- punishment and compensation measures should also be able to impose community services: if they are non-compliant, they must do something for the community.
- Putting supervision mechanisms in place to ensure the OPC is not, and is not becoming a partisan from lobbying groups, political parties or businesses: considerations as to designating the representative by a multi-party committee and not only the party in power.

**General conditions for testing process were identified as follows:**

- Provide coaching and measures within the company to ensure that there is meaningful awareness of potential social impacts in the design and scaling of an AI product, as this is the responsibility of the tech company.
- Including & pushing a disclaimer informing the user that an AI state is being used in the product, and not wait for the user to request such information.
- Leveraging Canadian diversity as a national strength, to ensure diverse feedback at the time of testing, from a vast array of the population.
- Creating a mandatory interactive training of the company with the user, to provide for a mandatory formation of the user by the company, when using product or service. (as an analogy, banks already a mandatory obligation to augment financial literacy; this can be done to and for tech businesses and users)
- Mandatory training of tech experts in ethics and vice versa.

**Topic 5.-Can the legal principles of purpose specification and data minimization work in AI (artificial intelligence) context and be designed for at the outset? If yes, would doing so limit potential societal benefit to be gained from use of AI (artificial intelligence)? If not, what are the alternatives or safeguards to consider?**

**Short Reply:**

Purpose specification and data minimization are insufficient in terms of mechanisms for dealing with AI systems, and in some cases are impractical.

**The group recommended:**

- Imposing respect of human rights explicitly in the legal framework for governance
- Identifying clear Go/No Go Zones, or Go/No Go Rules, for practical guidance.

**Comments:**

The group proposed that human rights be used as a "ceiling" for limiting specific AI applications within the general data protection and privacy framework. Such limitations are to include limitations on the use of data. The proposed test to protect individuals was identified as a rule: Data cannot be used to detriment anyone's life.

The group was also aware that there were limits as to the practicality of the consent model which became near infeasible in the context of AI systems evolving daily, and for which it is potentially impossible to know for which purpose it is achieving its tasks.

It was proposed that a list of things, activities or purposes which would be clearly and explicitly not permitted or unauthorized would be more effective, similarly to that of the Go/No Go zones. Despite limits of such identification, some purposes and some applications appear reasonably clear that there are permissible and not, and these should be highlighted for practical guidance. An example of such No Go Zone - or No Go Rule- is that data cannot be used to harm society, and should therefore not exacerbate differences or political uses.

**Topic 6.-Is it fair to consumers to create a system where, through the consent model, they would share the burden of authorizing AI (artificial intelligence) versus one where the law would accept that consent is often not practical and other forms of protection must be found?**

**Short Reply**:

Consent model should remain but should be augmented, while being supported by other forms of protection as well. A combined approach would be optimal.

The group recommended:

- Using plain language instead of "legalese" for terms and conditions for consent.
- Imposing mandatory opt-outs for "optional" data versus necessary data (data minimization), as well as imposing more flexible settings in terms of different options for use of data, instead of an "All-in" or "All-out" approach.

**Comments**:

Challenges linked with the consent model were identified as unclear, and very long and complex terms and conditions written in legalese. It was proposed that instead of catering to protect the liability of the business or corporation, the privacy policy and consent models used should be catered in a manner understandable and useful for the user's interests as well, since he has to provide consent in a meaningful manner.

It was proposed that the use of general language should be made for conditions applicable to the consent model or mechanism, as understandable by any user -and not as readable by lawyers only-, in order to provide for actual consent.

It was additionally proposed that in cases where consent may be absolutely impractical, access to data audits could be seen as an alternative method for consent mechanism to be withdrawn.

Finally, it was proposed that for further options to the user or individual concerned for the use of data should be available, or an opt-out opportunity, and that general guidelines should let a user know how data can or should be used even in the case of AI.

**Topic 7.-What could be the role of de-identification or other comparable state of the art techniques (synthetic data, differential privacy, etc.) in achieving both legitimate commercial interests and protection of privacy?**

**Short Reply**:

De-identification is necessary and should be held at the highest level of priority.

**Proposal:**

- Define tangible criteria to use to de-identify information
- Government should work with technical regulator to monitor non-compliance and judge upon new cases and non-compliant cases

**Comments**:

The group brought up the unclear scope of the concept of legitimate commercial interests and identified potential dilemmas between what needs to be used and what the business would want to use to better exploit its business. The question that arose was: what type of data would be considered legitimate, and which one would not?

The general agreement was that there should not be a stringent prohibition preventing the collection of data since it is not possible to know how the data could be use -for social and individual good- in the future.

Concerns were raised about AI systems and de-identification techniques being like "moving targets", rendering the imposition of a specific technique in regulation as difficult, but then which inevitably pushed businesses into using risk management models. It was said that de-identification may at times even lose all meaning since technologies are evolving and it is not possible to know in the long run what will be possible in terms of re-identification.

Despite the above, de-identification techniques and organizational measures, were seen as a high priority, and the group converged towards proposing a fixed security measure combined with a certain flexibility and judgement-based assessment or mechanism to monitor the practices. It was suggested that IT auditors would become a necessity to verify compliance with such security criteria.

One proposal was to have tiers of data, or different categories of data, which should be held with different security standards depending on the sensitivity of the data.

**Topic 8.-Is data traceability necessary, in an AI (artificial intelligence) context, to ensure compliance with principles of data accuracy, transparency, access and correction and accountability, or are there other effective ways to achieve meaningful compliance with these principles?**

**Short Reply**:

Yes, data traceability is necessary, and should be imposed as a mandatory business practice, but technical and organizational measures should be put in place, to ensure that such traceability records would or could not be used for re-identification of personal data.

However, it should be combined with other ways to effectively achieve compliance.
A few examples that were explored were:

- Creating a presumption, or explicitly shifting the burden of proof at law relative to compliance, non-negligence, respect of human rights and of privacy upon the business, and away from the individual.
- Mandatory internal and external audit mechanisms

**Comments**:

The question in the manner it was drafted was found to be ambiguous in meaning and in scope, and it was suggested it would be good for it to be more specific. For instance, the group had difficulty if the traceability related to the manner in which the data was input, processes, weighed and recorded within the AI systems, or if it referred to the manner and method of tracing back the data sources to their origins, as well as to whom received such data.

It was proposed that every step should be documented, including the metadata in an unaltered fashion, and it should be made easily ready for audit from a legitimate body. This would provide for additional protection pertaining to explainability, accountability and auditability. The need for guidelines and standards for documentation arose from the discussion.

Concerns were raised about how the concept of audit, from traditional accounting, was increasingly shifting, and the need to consult specific technical agencies was identified. The solution of auditing compliance was justified on grounds that audit from an independent and neutral third-party would treat everybody equally.

In conclusion, data traceability should be insured to the extent that data was de-identified before cannot become identifiable with tracing.

## SOURCES:

## ARTICLES:

## LEGAL

*EU Guidelines for Trustworthy AI,*
https://ec.europa.eu/digital-single-market/en/news/ethics-guidelines-trustworthy-ai

*Personal Information Protection and Electronic Documents Act*, S.C. 2000, c.-5

*Act Respecting the Protection of Personal Information in the Private Sector*, c. P-39.1

Office of the High Commissioner for Human Rights, UN Guiding Principles on Business and Human Rights, United Nations, 42 pages.

*Reference re Same-Sex Marriage*, [2004] 3 S.C.R. 698, 2004 SCC 79, "living tree"

*We have not added footnotes or references pertaining to such sources and references listed and identified by OPCC on Consultation proposals web page. Please refer to such page for additional references.*